\documentclass[11pt,twoside]{article}

\usepackage{graphicx} 
\usepackage{amsmath,amssymb} 

\def\bb1{\textup{\small{1}} \kern-3.6pt \textup{1}\ }

\DeclareFontFamily{U}{rsfs}{}         % Formal Script            %
\DeclareFontShape{U}{rsfs}{m}{n}{<5> rsfs5 <6><7> rsfs7          %
  <8><9><10><10.95><12><14.4><17.28><20.74><24.88> rsfs10}{}     %
\DeclareMathAlphabet{\mathfs}{U}{rsfs}{m}{n}                     %
\newcommand{\mfs}[1]{\mathfs {#1}}                               %
\newcommand{\sH}{{\mfs H}}

\begin{document}
\thispagestyle{empty}  
\vspace*{-1cm} 
\hspace*{\fill}\parbox{3.2cm}{{\small MPI--PhT--99/46} \newline 
	{\small LMU--TPW--99/18}}
\begin{center} 
\vspace*{2 truecm} 
\Large\bf Loop Quantum Gravity and the Meaning of Diffeomorphism Invariance \\[1.2 truecm] 
\normalsize \rm
\renewcommand{\thefootnote}{\fnsymbol{footnote}}
Marcus Gaul${}^{1,3,4,}$\footnote[2]{\ttfamily{mred@mppmu.mpg.de}}
and Carlo Rovelli${}^{1,2,}$\footnote[3]{\ttfamily{rovelli@pitt.edu}}\\[0.8 truecm]
\renewcommand{\thefootnote}{\arabic{footnote}}
{\small\it 
${}^1$Departement of Physics and Astronomy, University of Pittsburgh, Pittsburgh,\\ 
PA 15260, USA\\[0.1 truecm]
${}^2$Centre de Physique Th\'eorique, CNRS Luminy, F-13288 Marseille, France\\[0.1 truecm]
${}^3$Max--Planck--Institut f\"ur Physik, F\"ohringer Ring 6, 
D-80805 M\"unchen, Germany\\[0.1 truecm]
${}^4$Sektion Physik, Ludwig--Maximilians--Universit\"at, 
Theresienstr.\ 37,\\ D-80333 M\"unchen, Germany}\\
\vspace{0.9cm}
{\small December 20, 1999}\\
\vspace{0.9cm}
{\bf Abstract}\\
\end{center} 
{\small
This series of lectures presented at the {\it 35th Karpacz Winter School 
on Theoretical Physics: From Cosmology to Quantum Gravity} gives a simple and
self-contained
introduction to the non-per\-tur\-ba\-tive and background independent loop 
approach of canonical quantum gravity.
The Hilbert space of kinematical quantum states is constructed and a complete basis
of \emph{spin network states} is introduced.
An application of the formalism is provided by the spectral analysis
of the area operator, which is the quantum analogue of the classical 
area function. 
This leads to one of the key results of loop quantum gravity:
the derivation of the discreteness of the geometry and the computation of the quanta of area.
Finally, an outlock on a possible covariant formulation of the theory is given
leading to a \lq \lq sum over histories'' approach, denoted as \emph{spin foam model}.
Throughout the whole lecture great significance is attached to conceptual and 
interpretational issues. In particular, special emphasis is given to the role played 
by the diffeomorphism group and the notion of observability in general relativity.}

\newpage
\vspace*{0.5cm}
\tableofcontents
\newpage

%%%%%%%%%%%%%%%%%%%%%%%%%%%%%%%%%%%%%%%%%%%%%%%%%%%%%%%%%%%%%%%%%%%%%%%%%%%%%%%%%%%%%%%
%%%%%%%%%                      Introduction (Section 1)                       &&&&&&&&&
%%%%%%%%%%%%%%%%%%%%%%%%%%%%%%%%%%%%%%%%%%%%%%%%%%%%%%%%%%%%%%%%%%%%%%%%%%%%%%%%%%%%%%%

\section{Introduction}
In the beginning of this century, physics has undergone two great conceptual
changes.
With the discovery of general relativity and quantum mechanics 
the notions of matter, causality, space and time experienced the biggest modifications
since the age of Descartes, Copernicus, and Newton. 
However, no fully convincing synthesis of these theories exists so far. Simple
dimensional analysis reveals that new predictions of a quantum theory of gravitation
are expected to take place at the Planck length 
$l_P \equiv \left( \hbar G / c^3 \right)^{1/2} \sim 10^{-35}\,\mbox{m}$.
This scale appears to be far below any current 
experimental technique. Nevertheless, quite recently interesting proposals 
and ideas to probe experimentally the physics at the Planck scale have been suggested
\cite{AmelinoCamelia99,Ellis99}. 

From the theoretical point of view, several approaches to a theory 
of quantum gravity have emerged, inspired by various research fields in contemporary 
physics and mathematics.
The most popular research direction is string theory, followed by loop quantum 
gravity. Other directions range from discrete methods to non-commutative geometry. 
We have listed the main current approaches to a quantum theory of gravity (which are, 
by the way, far from being independent) in Table \ref{approaches_to_QGR}.
Despite this variety of ideas and the effort put in so far, many
questions are still open. For an overview and a critical comparison of the different
approaches, see \cite{Rovelli97b}.

\vspace*{0.3cm}
% The following slightly changed table was taken from Carlo's paper gr-qc/9803024
% ``Strings, loops and others...''
\begin{table}[h] 
\scriptsize
\begin{tabular}{@{\hspace{5mm}}ccc}
& & \\
& & \\
\normalsize\bf Traditional & \normalsize\bf Most Popular & \normalsize\bf New  \\
& & \\
& & \\
        \framebox{
           \begin{minipage}{110pt}
               {\bf Discrete methods} \\ 
              \mbox{\ } Dynamical triangulations\\
              \mbox{\ } Regge calculus\\ 
              \mbox{\ } Simplicial models
        \end{minipage}} 
& 
    \framebox{
        \begin{minipage}{117pt} \mbox{\ } \\
             {\bf String theory} \\ 
             \mbox{\ }$\rightarrow$ {\em Black hole entropy} \\ 
         \end{minipage}} 
& 
      \framebox{
         \begin{minipage}{123pt} \mbox{\ } \\
              {\bf Non-commutative geometry} \vspace*{1.5mm} 
          \end{minipage}}
\\ && \\
         \framebox{
              \begin{minipage}{110pt}
                    {\bf Approximate theories} \\
                       \mbox{\ } Euclidean quantum gravity \\
                       \mbox{\ } Perturbative quantum gr. \\
                       \mbox{\ } QFT on curved space--times
               \end{minipage}}
&  
         \framebox{ 
               \begin{minipage}{117pt} \mbox{\ } \\ 
                    {\bf Loop quantum gravity} \\  
                     \mbox{\ } $\rightarrow$ {\em Black hole entropy}\\
                     \mbox{\ } $\rightarrow$ {\em Eigenvalues of geometry:}\\ 
                     \\ 
                     \mbox{} \fbox{ 
                     $\! A_{\vec j} = 8 \pi\hbar G 
                     \sum_{i}\sqrt{j_{i}(j_{i}+1)}$}\\
                \end{minipage}} 
& 
          \framebox{
         \begin{minipage}{123pt} \mbox{\ } \\
              {\bf Null surfaces} \vspace*{1.5mm} 
          \end{minipage}} 
\\
& &  \framebox{
           \begin{minipage}{123pt}  \vspace*{.5mm}
              {\bf Spin foam  models}   \\
              \mbox{\ } {\em $\rightarrow$ convergence of loop, \\
              \mbox{\ \ \ \ \ } discrete, TQFT \\ 
              \mbox{\ \ \ \ \ } and sum-over-histories}  \vspace*{.5mm} 
          \end{minipage}} \\ 
      \framebox{
           \begin{minipage}{110pt}{\bf Unorthodox approaches} \\  
                \mbox{\ }  Sorkin's Posets\\  
                \mbox{\ }  Finkelstein\\  
                \mbox{\ }  Twistors\\ 
                \mbox{\ } \ldots \vspace*{.5mm}
            \end{minipage}} 
&  &   \\ 
&  &   \\ 
\end{tabular}
\begin{flushleft}
\parbox{9.4cm}{\caption{The main current approaches to quantum gravity.}}
\end{flushleft}
\label{approaches_to_QGR}
\end{table}

\newpage
String theory was inspired and constructed mainly by particle physicists.
Its attitude towards the fundamental forces is to treat general relativity on an equal 
footing with the field theories describing the other interactions, the destinctive 
feature being the energy scale. String theory is supposed to be a theory 
of all interactions---electromagnetic, strong, weak and gravitational---
which are treated in a unified quantum framework. Classical 
(super-)gravity emerges perturbatively as a low-energy limit in superstring theory.
Until 1995 the problem was the lack of a non-perturbative formulation 
of the theory. This situation has improved with the discovery of string dualities, 
\lq \lq D-branes'', and \lq \lq M-theory'' in the so-called 
{\it 2nd superstring revolution}. 
Nevertheless, despite the recent exciting discoveries in M-theory and the AdS/CFT 
equivalence, a complete non-perturbative or strong-coupling formulation of 
string/M-theory is still not in sight.

A point that is often criticized in string theory by relativists is
the lack of a background independent formulation, i.e. invariance under 
active diffeomorphisms, which is one of the fundamental principles of 
general relativity.
String/M-theory is formulated on a (implicitly) fixed background geometry which is 
itself not dynamical. In a truely background independent formulation, no reference
to any classical metric should enter neither the definition of the state space 
nor the dynamical variables of the theory.
Rather the metric should appear as an operator allowing for quantum 
states which may themselves be superpositions of different backgrounds.

In fact, relativists do not view general relativity as an 
additional item in the list of the field theories describing 
fundamental forces, but rather as a major change in the manner 
space and time are described in physics.  This point is often 
misunderstood, and is often a source of confusion; it might be 
worthwhile spending a few additional words.  The key point is not 
that the gravitational force, by itself, must necessarilly be 
seen as different from the other forces: the point of view that 
the gravitational force is just one (and the weakest) among 
the interactions is certainly viable and valuable.  Rather, the 
key point is that, with general relativity, we have understood 
that the world is {\em not\/} a non-dynamical metric manifold 
with dynamical fields living over it.  Rather, it is a collection 
of dynamical fields living, so to say, in top of each other.  The 
gravitational field can be seen---if one wishes so---as one among 
the fields. But the definiton of the theory over a given 
background is, from a fundamental point of view, physically 
incorrect.

Loop quantum gravity is a background independent approach to quantum gravity.
For many details on this approach, and for complete references, see \cite{Rovelli97a}.
Loop quantum gravity has been developed \emph{ab initio} as a
non-per\-tur\-ba\-tive and background independent canonical quantum theory of gravity.
Besides ordinary general relativity and quantum mechanics no additional input 
is needed.
The approach makes use of the reformulation of general relativity as a 
dynamical theory of connections. 
Due to this choice of variables the phase space of the theory resembles at the 
kinematical level closely that of conventional $SU(2)$ Yang--Mills theory. 
The main ingredient of the appraoch is the choice of holonomies of 
the connections---the loop variables---as the fundamental degrees of freedom
of quantum gravity. 

The philosophy behind this approch is different from string theory as one
considers here standard 4-dimensional general relativity trying to develop a
theory of quantum gravity in its proper meaning without claiming to describe a unified 
picture of all interactions. 
Loop quantum gravity is successful in describing Planck-scale phenomena.
The main open problem, on the other hand, is the connection
with low-energy phenomena. In this respect loop quantum gravity has opposite strength
and weakness than string theory.
However, some of the conceptually different 
approaches that were given in Table~\ref{approaches_to_QGR} show surprising similarities which 
could be a focal point of attention 
for the future \footnote{\label{simresults}For instance, 
string/M-theory, non-commutative geometry and loop quantum gravity 
seem to point to a similar \emph{discrete} short distance space--time structure.  
Suggestions have been made that a complete theory must involve elements 
from each of the approaches. 
For further details we refer to \cite{Smolin_Karpacz99,Smolin97}.}.

One might wonder how one can hope to have a consistent non-perturbative formulation 
of quantum gravity when perturbative quantization of covariant
general relativity is non-renormalizable. However, the basic assumption in 
proving the non-renormalizability of general relativity is the availability of 
a Minkowskian space--time at arbitrarily short distances, 
an assumption which is certainly not correct in a theory of quantized  
gravity, i.e. in a quantum space--time regime.
As will be discussed later, one of the key results obtained so far in loop quantum gravity 
has been the calculation of the quanta of geometry \cite{RovelliSmolin95}, 
i.e. the spectra of the quantum analogues to the classical area and volume functionals.
Remarkably they turned out to be discrete! This result (among similar ones obtained in 
other approaches, see footnote~\ref{simresults}) 
indicates the existence of a quantum space--time structure at the Planck scale 
which doesn't have to be continuous anymore. More specifically this implies the 
emergence of a natural cut-off in quantum gravity that might also account as a regulator 
of the ultraviolet divergencies plaguing the standard model.
Thus, standard perturbative techniques in field theory cannot be taken for granted
at scales where quantum effects of gravity are expected to dominate.

General relativity is a constrained theory. Classically, the constraints are equivalent
to the dynamical equations of motion. The transition to the quantum theory is carried
out using canonical quantization by appliying the algorithm developed by Dirac 
\cite{Dirac64}.
In the loop approach, the unconstrained classical theory is quantized, requiring
the implementation of quantum 
constraint operators afterwards. Despite many results obtained in the last few years, 
a complete implementation of all constraints including the Hamiltonian constraint, 
which is the generator of \lq \lq time evolution'', i.e. the dynamical part of the theory, 
is still elusive. This is of course not surprising, since we do not expect to be able to obtain
a complete solution of a highly non-trivial and non-linear theory.
To address this issue, covariant methods to understand the dynamics have been developed
in the last few years. These can be obtained from a
\lq \lq sum over histories'' approach, derived from the canonical formulation. This
development has led to the so-called \emph{spin foam models}, in which
spin networks are loosely speaking \lq \lq propagated in time'', leading to a 
space--time formulation of loop quantum gravity. This formulation of the theory provides
a starting point for approximations, offers a more intuitive understanding of quantum 
space--time, and is much closer to particle physics methods.
A brief description will be given below in sect.~\ref{sect_hamiltonian_constraint}.

These lectures are organized as follows. 
We start in sect.~\ref{sect_basics} with the basic mathematical framework of loop quantum 
gravity and end up with the definition of the kinematical Hilbert space of quantum gravity.
In the next section an application of these tools is provided by constructing the basic 
operators on this Hilbert space. We calculate in a simple manner the spectrum of what 
is going to be physically interpreted as the area operator.
Section \ref{sect_diffeoInv} deals with the important question of observability in 
classical and quantum gravity, a topic which is far from being trivial, and the 
meaning of diffeomorphism invariance in this context.
In the end of theses notes, we will give the prospects for a dynamical 
description of loop quantum gravity, which is encoded in the concept of 
\emph{spin foam models}. One such ansatz is briefly discussed in sect. \ref{sect_dynamics}.
The following final section concludes with future perspectives and open problems.

For the sake of completeness, we give in Table 2 a short historical 
survey of the main achievements in canonical quantum gravity since the  
reformulation of general relativity in terms of connection variables.
A more detailled discussion of some of these aspects and full bibliography is given
in \cite{Rovelli97a}.
\begin{table}
\renewcommand{\arraystretch}{1.5}
\setlength\tabcolsep{10pt}
\begin{center}
\begin{tabular}{p{1cm}p{6cm}p{2cm}}
\noalign{\smallskip}
\hline
\noalign{\smallskip}
'86 & Classical Connection Variables & \cite{Sen82,Ashtekar86,Ashtekar87} \\
'87 & Lattice Loop States solve $\hat{H}$ & \cite{JacobsonSmolin88}\\
'88 & Loop Quantum Gravity & \cite{RovelliSmolin88,RovelliSmolin90}\\
'92 & Weave States & \cite{AshtekarRovelliSmolin92}\\
'92 & Diffeomorphism Invariant Measure & \cite{AshtekarIsham92,Baez93}\\
'95 & Spin Network States & \cite{RovelliSmolin95b}\\
'95 & Volume and Area Operators & \cite{RovelliSmolin95}\\
'95 & Functional Calculus & \cite{AshtekarLewand95,Ashtekaretal95}\\
'96 & Hamiltonian Operator & \cite{Thiemann96,Thiemann96b}\\
'96 & Black Hole Entropy & \cite{Rovelli96,Krasnov97,AshBaezCorKras98}\\
'98 & Spin Foam Formulation & \cite{ReisenbergerRovelli97,Baez98}\\
\noalign{\smallskip}
\hline
\noalign{\smallskip}
\end{tabular}
\end{center}
\caption{A short historical survey of canonical quantum gravity, with some relevant references.}
\label{history}
\end{table}

%%%%%%%%%%%%%%%%%%%%%%%%%%%%%%%%%%%%%%%%%%%%%%%%%%%%%%%%%%%%%%%%%%%%%%%%%%%%%%%%%%%%%%%
%%%%%%%%%                              Section 2                              &&&&&&&&&
%%%%%%%%%%%%%%%%%%%%%%%%%%%%%%%%%%%%%%%%%%%%%%%%%%%%%%%%%%%%%%%%%%%%%%%%%%%%%%%%%%%%%%%

\section{The Basic Formalism of Loop Quantum Gravity}
\label{sect_basics}
Our attention in this lecture will be focused on conceptual foundations and the 
development of the main ideas behind loop quantum gravity. 
However, because of the highly mathematical nature of the subject some 
technical details are unavoidable, thus this section is devoted 
to the essential mathematical foundations.

The reader is not assumed to be familiar with the connection variables,
which constitute the basis for most efforts in canonical quantum gravity since  
1986. Thus we start by considering the canonical formalism in the connection approach,
which is reviewed in \cite{Rovelli91}.
For a recent overview of loop quantum gravity and a comprehensive list of 
references we refer to \cite{Rovelli97a}.

\subsection{A brief Outline of the Connection Formalism}
\label{sect_connectionformalism}
In loop quantum gravity, we construct the quantum theory using canonical quantization.
This is analogous to ordinary field theory in the functional Schr\"odinger representation.
The approach may be called conservative in the sense that originally no new 
structures like supersymmetry\footnote{Nevertheless, there exist extensions
of the Ashtekar variables to supergravity \cite{Jacobson88}, and quite recently 
$N=1$ supersymmetry was introduced in the context of spin networks \cite{LingSmolin99}.},
extended objects, or extra dimensions are postulated.
(It is important to emphasize, in this context, the fact, 
sometimes forgotten these days, that supersymmetry, 
extended objects or extra dimensions are interesting theoretical 
hypotheses, not established properties of Nature!).  The approach 
aims at unifying quantum me\-cha\-nics and general relativity by developing 
new non-perturbative techniques from the outset and by staying as close as possible to 
the conventional settings of quantum theory and experimentally tested general relativity.

The foundations of the formalism date back to the early 60s 
when the \lq \lq old'' Hamiltonian or canonical formulation of classical general 
relativity, known as ADM formalism, was constructed.
The canonical scheme is based on the construction of the phase space $\Gamma$. 
Phase space is a covariant notion. It is 
the space of solution of the equations of motion, modulo 
gauges. However, in order to coordinatize $\Gamma$
explicitly, one usually breakes explicitly covariance and splits 
4-dimensional space--time $\mathcal{M}$ into 3-dimensional space 
plus time. We insist on the fact that this breaking of 
covariance is not structurally needed in order to set up the 
canonical formalism; rather it is an artefact of the 
coordinatization we choose for the phase space. We take
$\mathcal{M}$ to have topology $\mathbb{R} \times M$, 
and we cover it with a foliation $M_t$. Here $M$ is the 3-manifold 
representing \lq \lq space'' and $t \in \mathbb{R}$ is a 
(unphysical) time parameter.  The basic variables on phase space 
are taken to be the induced 3-metric $q_{ab}(x)$ on $M$ and the 
extrinsic curvature $K_{ab}$ of $M$.

The easiest construction of the connection variables is given by first 
reformulating the 
ADM--formalism of canonical gravity in terms of (local) triads $e^i_a(x)$, which satisfy 
$q_{ab}(x) = e^i_a(x) e^i_b(x)$. This introduces an additional local $SU(2)$ 
gauge symmetry into the theory, which geometrically corresponds to arbitrary local 
frame rotations.
One obtains $\big(E^a_i(x),K^i_a(x)\big)$ as the new canonical pair on phase space 
$\Gamma$. $E^a_i$ is the \emph{inverse} 
densitized triad\footnote{In the literature one often finds
tensor densities marked with an upper tilde for each 
positive density weight and 
a lower tilde for each negative weight, such that the densitized triad is often written as
\mbox{$\widetilde{E}^i_a(x) := e(x) \, e^i_a(x)$}.}, i.e. a vector with 
respect to $SU(2)$ and density weight one. The densitized triad itself is defined
by $E^i_a := e \, e^i_a$, where $e$ is the determinant of $e^i_a$. 
The indices \mbox{$a,b, \ldots = 1,2,3$} 
refer to spatial tangent space components, while $i,j, \ldots = 1,2,3$ are internal 
indices that can be viewed as labelling either the axis of a local triad or the basis of 
the Lie algebra of $SU(2)$.
The inverse triad $E^a_i$ is the square-root of the 3-metric in the sense that 
\begin{equation} 
  \label{metric_to_Ashtekar}
  E^a_i(x) E^b_i(x) = q(x) \, q^{ab}(x) \;,
\end{equation}
where $q(x)$ is the determinant of the 3-metric $q_{ab}(x)$. 
The canonically conjugate variable $K^i_a(x)$ of $E^a_i(x)$
is again closely related to the extrinsic curvature 
of $M$ via \mbox{$K^i_a= K_{ab} E^{bi}\!/ \! \sqrt{q}$}.

The transition to the connection variables is made
using a canonical transformation on the (real) phase space,
\begin{equation} 
\label{Ashtekar_connection}
A^i_a(x) = \Gamma^i_a(x) + \beta \, K^i_a(x) \;.
\end{equation}
Here $\Gamma^i_a(x)$ is the $SU(2)$ spin connection compatible with the triad,
and $\beta$, the Immirzi parameter, is an arbitrary real constant.
The original Ashtekar--Sen connection $A(x)$ was introduced in 1982 as a complex 
self-dual connection on the spatial 3-manifold $M$, corresponding
to $\beta=i$ in (\ref{Ashtekar_connection}). Here we will use the real 
formulation.

Nevertheless $\big(A^i_a(x), \, E^a_i(x)\big) \in \Gamma$ form 
a canonical pair on the phase space $\Gamma$ of general relativity \cite{Ashtekar86,Ashtekar87}. 
Here $A^i_a$ has to be considered 
as the new configuration variable, while the inverse densitized triad $E^a_i$ corresponds 
to the canonically conjugate momentum.
With this reformulation classical general relativity 
has the same kinematical phase space structure as an $SU(2)$ Yang--Mills theory.

The Poisson algebra generated by the new variables is 
\begin{eqnarray}
  \big\{ E^a_i(x), \, E^b_j(y) \big\} &=& 0 \;, \hspace{0.5cm}
  \big\{ A^i_a(x), \, A^j_b(y) \big\} = 0 \;, \nonumber \\
  \label{Poissonalgebra}
  \big\{ A^i_a(x), \, E^b_j(y) \big\} &=& 
  \beta \, G \, \delta^i_j \, \delta^b_a \, \delta^3 (x,y) \;, 
\end{eqnarray}
where $G$ is the usual gravitational constant. It arises because the 
conjugate momentum of the configuration variable $A^i_a$ (obtained as the derivative
of the Lagrangian with respect to the velocities) is actually given by $1/G \times E^a_i$.
As a consequence of the 4-dimensional diffeomorphism invariance of general relativity,
the (canonical) Hamiltonian vanishes weakly\footnote{This is indeed true for 
\emph{any} generally covariant theory, which means that
a theory whose gauge group contains the diffeomorphism group of 
the underlying manifold has a weakly vanishing Hamiltonian. Here
weakly vanishing refers to vanishing on physical configurations.}.
The full dynamics of the 
theory is encoded in so-called first-class constraints which are functions on 
phase space that vanish for physical configurations. 
The constraints generate transformations between those classical configurations that 
are physically indistinguishable.
The first-class constraints of canonical general relativity are the familiar 
Gauss law of Yang--Mills theory, which generates local $SU(2)$ gauge transformations, 
the diffeomorphism constraint generating 3-dimensional diffeomorphisms 
of the 3-manifold $M$, and the Hamiltonian constraint, which is 
the generator of the evolution of the inital spatial slice $M$ in 
coordinate time. The Gauss constraint enters the theory as a result of the 
choice of triads and it makes general relativity resemble a 
Yang--Mills gauge theory. And indeed, the phase spaces of both theories are
similar. The constrained surface of general relativity is embedded in 
that of Yang--Mills theory apart from the additional local restrictions which 
appear in gravity besides the Gauss law. 
 
The use of the set of canonical variables involving a
\emph{complex} connection $A(x)$ leads to a simplification of the 
Hamiltonian constraint.
With the use of a \emph{real} connection, the constraint loses its simple 
polynomial form. At first, this was considered as a serious obstacle for 
quantization. However, Thiemann succeeded in constructing a 
Lorentzian quantum Hamiltonian constraint \cite{Thiemann96b} 
in spite of the non-polynomiality of the classical expression. His work has prompted
the wide use of the real connection, a use which was first advocated by 
Barbero \cite{Barbero95}.

We will now briefly describe the quantum implementation of this kinematical setting.
The canonical variables $A$ and $E$ (or functions of these), are replaced 
by quantum operators acting on a Hilbert space of
states, promoting Poisson brackets to commutators.
In other words, an algebra of observables should act on a Hilbert space.  
More precisely, we establishe an isomorphism between the Poisson algebra of 
classical variables and the algebra generated by the corresponding 
Hermitian operators by introducing a linear operator representation
of this Poisson algebra. 
The quantum states are normalizable functionals over 
configuration space, i.e. functionals of the connection $\Psi(A)$ (or limits of these).
The subset of physical states is obtained 
from the set of all wavefunctions on $M$ by imposing the
quantum analogues of the contraints, i.e. by requiring
the physical states to lie in the kernel of all quantum constraint 
operators\footnote{A distinct quantization method is the 
\emph{reduced phase space quantization},
where the physical phase space is constructed classically by solving 
the constraints and factoring out gauge equivalence prior to quantization.
But for a theory as complicated as general relativity it seems impossible to 
construct the reduced phase space.  The two methods could lead to inequivalent 
quantum theories.  Of course, it is possible, in principle, that 
more than one consistent quantum theory having general relativity 
as its classical limit might exist.}.

The space of physical states must have the structure of a Hilbert 
space, namely a scalar product, in order to be able to compute 
expectation values.  This Hilbert structure is determined by the 
requirement that {\em real\/} physical observables correspond to 
{\em self-adjoint\/} operators.  In order to define a Hilbert 
structure on the space of {\em physical\/} states, it is 
convenient (althought not strictly necessary) to define first a 
Hilbert structure on the space of {\em unconstrained\/} 
states.  This is because we have a much better knowledge of the 
unconstrained observables than of the physical ones.  If we 
choose a scalar product on the unconstrained state space which is 
gauge invariant, then there exist standard techniques to ``bring 
it down'' to the space of the physical states.  Thus, we need a 
gauge and diffeomorphism invariant scalar product, with respect 
to which real observables are self-adjoint operators. 

\subsection{Basic Definitions}
In this subsection we start with the actual topic of the lecture, the construction
of \emph{loop quantum gravity}.
Space--time is assumed to be a 4-dimensional Lorentzian manifold $\mathcal{M}$
with topology $\mathbb{R} \times M$, where $M$ is a
real analytic and orientable 3-manifold. For simplicity we take $M$ to be topologically
$S^3$. Loosely speaking $M$ represents \lq \lq space'' 
while $\mathbb{R}$ refers to \lq \lq time''.

On $M$ we define a smooth, Lie algebra valued connection 1-form $A$, i.e. \linebreak
\mbox{$A(x)=A^i_a(x)\,\tau_i\,dx^a$}, where $x$ are local coordinates on $M$, 
$A^i_a(x) \in C^{\infty}(M)$, and \linebreak
\mbox{$\tau_i= (i/2)\, \sigma_i$} are the $SU(2)$ generators
in the fundamental representation, satisfying \linebreak
\mbox{$[\tau_i, \tau_j] = \epsilon_{ijk} \tau^k$}.
Here $\sigma_i$ are the Pauli matrices.
The indices $a,b,c=1,2,3$ play the role of tangent space indices while $i,j,k=1,2,3$ 
are abstract internal $su(2):=Lie\big(SU(2)\big)$ indices\footnote{One may consider 
a principal $G$-bundle $P$ over $M$, with 
structural group $G=SU(2)$ (i.e. compact and connected). 
The classical configuration space $\mathcal{A}$ of general relativity is given 
by the smooth connections on $P$ over $M$.
The principal $SU(2)$-bundles are in this case topologically trivial, 
hence the $SU(2)$ connections
on $P$ can be represented by $su(2)$-valued 1-forms, since a global cross-section
can be used to pull back the connections to 1-forms on $M$.}.
We call $\mathcal{A} = \{ A \}$ the space of smooth connections on $M$ 
and denote continuous functionals on $\mathcal{A}$ as $\Psi(A)$.
These functionals build up a topological vector space $L$ under the pointwise topology.

\subsection{The Construction of a Hilbert Space $\mathcal{H}$}
\label{sect_Hilbertspace}
In order to define a Hilbert space $\mathcal{H}$ based on the above linear space $L$
of quantum states $\Psi(A)$,
an inner product needs to be introduced, i.e. an appropriate measure on the space 
of quantum states is required.
For that purpose the appearance of the \emph{compact} gauge group 
$SU(2)$ turns out to be essential. 
We demand the following properties of $\mathcal{H}$:
\begin{itemize}
\item $\mathcal{H}$ should carry a unitary representation of $SU(2)$
\item $\mathcal{H}$ should carry a unitary representation of $Diff (M)$ .
\end{itemize}

We construct the inner product by means of a special class of functions of 
the connection in $L$, the \emph{cylindrical functions}. For their 
construction we need some tools, namely \emph{holonomies} and \emph{graphs}.

\subsubsection{Holonomies.} 
\label{sect_Holonomies}
Let a curve $\gamma$ be defined as a continuous, piecewise smooth map from the intervall
$[0,1]$ into the 3-manifold $M$,
\begin{eqnarray}
\gamma : [0,1] &\longrightarrow& M \\
s &\longmapsto& \{\gamma^a(s)\} \;,\; a = 1,2,3 \;.
\end{eqnarray}
The holonomy or parallel propagator $U[A,\gamma]$, respectively, of the connection $A$ along 
the curve $\gamma$ 
is defined by
\begin{eqnarray}
U[A,\gamma](s) &\in& SU(2)\;, \\
U[A,\gamma](0) &=& \bb1 \;, \\
\label{paralleltransp}
\frac{d}{ds} \, U[A,\gamma](s) &+& A_a\big(\gamma(s)\big) \, \dot{\gamma}^a(s) 
        \, U[A,\gamma](s) =0 \;, 
\end{eqnarray}
where $\dot{\gamma}(s) := \frac{d \gamma(s)}{ds}$ is the tangent to the curve.
The formal solution of (\ref{paralleltransp}) is given by
\begin{equation}
U[A,\gamma](s) = 
        \mathcal{P} \exp \int_{\gamma} ds \, \dot{\gamma}^a \, A_a^i\big(\gamma(s)\big) 
        \, \tau_i \equiv \mathcal{P} \exp \int_\gamma A \;,   
\label{holonomy}
\end{equation}
in such a way that for any matrix-valued function $A\big(\gamma(s)\big)$ which is 
defined along $\gamma$, the path ordered expression (\ref{holonomy})
is given in terms of the power series expansion
\begin{eqnarray}
\label{pathordering}
\lefteqn{\mathcal{P} \exp \int_0^1 ds \, A\big(\gamma(s)\big)} \hspace{0.5cm} \nonumber \\
 & & = \sum_{n=0}^{\infty} \, \int_0^1 ds_1 \int_0^{s_1} ds_2 \, \cdots \int_0^{s_{n-1}} ds_n
	\, A\big(\gamma(s_n)\big) \cdots A\big(\gamma(s_1)\big) \;.
\end{eqnarray}
Here $\mathcal{P}$ denotes path ordering, i.e. the parameters $s_i$ are ordered with respect
to their moduli from the left to the right, or more explicitely $s_1 \leq s_2 \leq \ldots$ .

In a later section we will focus our attention to Wilson loops, which are
traces of the holonomy of $A$ along a curve $\gamma$, satisfiying $\gamma(0)=\gamma(1)$, 
i.e. a closed curve or loop, respectively, which in the following will be denoted as $\alpha$.
We write
\begin{equation}
T[A, \alpha] = - \mbox{Tr} \, U[A,\alpha] \; ;
\label{ta}
\end{equation}
these are gauge invariant functionals of the connection.

The key successful idea of the loop approach to quantum gravity 
\cite{RovelliSmolin90} is to choose the loop states
\begin{equation}
\Psi_{\alpha}(A) = - \mbox{Tr} \, U[A,\alpha] 
\label{ls}
\end{equation}
as the basis states for quantum gravity\footnote{\label{minus}The minus sign in (\ref{ta}) and 
(\ref{ls}) simplifies sign complications in the definition of the spin network states 
to be introduced later, see \cite{DePietriRovelli96}.}.
They are extended to disconnected loops, or multiloops\footnote{A multiloop 
is a collection of a finite number of loops $\{\alpha_{1}, \ldots, \alpha_{n}\}$,
which is, following tradition, also denoted as $\alpha$.},
respectively, by defining a multiloop state as $\Psi_{\alpha}(A) = \prod_{i}\left(- \mbox{Tr} \, 
U[A,\alpha_{i}]\right)$.  They have a number of remarkable 
features: they allow us to control completely the solution of the 
diffeomorphism constraint, and they \lq \lq largely'' solve the 
Hamiltonian constraint, as we will see later.  In QCD, states of 
this kind are unphysical, because they have infinite norm (they 
are \lq \lq too concentrated'', or \lq \lq not sufficiently smeared''). If 
in QCD we artificially declare these states to have finite norm, 
we end up with an unphysically huge, non-separable Hilbert space.  
In gravity, on the other hand, these states, or, more precisely, 
the equivalence classes of these states under diffeomorphisms, 
define finite norm states.  They are not too concentrated since 
in a sense they are---by diffeomorphism invariance---\lq \lq smeared all 
over the manifold''. Thus, they provide a natural and physical way to 
represent quantum excitations of the gravitational field.

For some time, however, a technical difficulty for loop quantum gravity 
was given by the fact that the states (\ref{ls}) form an overcomplete basis.  
The problem was solved in \cite{RovelliSmolin95b} by introducing the 
spin network states, which 
are combinations of loop states that form a genuine 
(non-overcomplete) basis. These spin network states will be constructed in the sequel.

\subsubsection{Graphs.}
\begin{figure}
\centerline{\includegraphics[width=6cm]{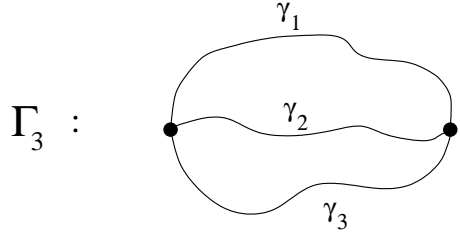}}
\caption[]{A simple example of a graph.}
\label{theta1}
\end{figure}

A graph \mbox{$\Gamma_n=\{\gamma_1,\ldots,\gamma_n\}$}
is a finite collection of $n$ (oriented) piecewise smooth curves or edges
$\gamma_i, \, i=1,\ldots,n$, respectively, embedded in 
the 3-manifold $M$, that meet, if at all, only at their endpoints.
As an example, consider the graph $\Gamma_3$ in Fig.~\ref{theta1} which is 
composed of three curves $\gamma_i$, denoted as \emph{links}.

\subsubsection{Cylindrical Functions.}
\label{subsect_cylfct}
Now pick a graph $\Gamma_n$ as defined above.
For each of the $n$ links $\gamma_i$ of $\Gamma_n$ consider the ho\-lo\-no\-my 
$U_i(A) \equiv U[A,\gamma_i]$ of the connection $A$ along $\gamma_i$. Every (smooth)
connection assigns a group element $g_i \in SU(2)$ to each link $\gamma_i$ of $\Gamma_n$ 
via the holonomy, \linebreak
$g_i \equiv U_i(A) = \mathcal{P} \exp \int_{\gamma_i} A$.
Thus an element of $[SU(2)]^n$ is assigned to the graph $\Gamma_n$.
The next step is to consider complex-valued functions 
$f_n(g_1, \ldots, g_n)$ on $[SU(2)]^n$,
\begin{equation}
f_n : [SU(2)]^n \rightarrow \mathbb{C} \;.
\end{equation}
These functions are Haar-integrable by construction, i.e.
finite with respect to the Haar measure of $[SU(2)]^n$ which is induced by that
of $SU(2)$ as a natural extension in terms of products of copies of it.

Given any graph $\Gamma_n$ and a function $f_n$, we define
\begin{equation}
\label{quantumstates}
\Psi_{\Gamma_n,f_n}(A) := f_n(U_1, \ldots, U_n)\;.
\end{equation}
These functionals depend on the connection only via the holonomies. 
They are called \emph{cylindrical functions}\footnote{The name 
\emph{cylindrical function} stems from integration theory on infinite-dimensional
manifolds, where they are introduced to define cylindrical measures. 
One can view the cylindrical function associated to
a given graph as being constant with respect to some (in fact, most) of the 
dimensions of the space of connections, i.e. as a cylinder on that space.}.
They form a dense subset of states in $L$, 
the space of continuous smooth functions on $\mathcal{A}$, defined above.
This justifies the exclusive use of this special class of functions for the 
construction of the Hilbert space.

As an example, we consider the cylindrical function corresponding to the graph 
$\Gamma_3$ in Fig.~\ref{theta1}.
Let $f_3$ be defined as 
\begin{equation}
f_3(U_1, U_2, U_3) := \mbox{Tr}(U_1 U_2 U_3) \;.
\end{equation}
Hence it follows that the cylindrical function corresponding to $\Gamma_3$
is given by
\begin{equation}
\Psi_{\Gamma_3,f_3}(A) = \mbox{Tr}\left(U[A,\gamma_1] U[A,\gamma_2] U[A,\gamma_3]\right) 
\;.
\end{equation}
An important property of cylindrical functions which turns out to be essential for the
definition of an inner product is the following. A 
cylindrical function based on a graph $\Gamma_n$ can always be rewritten as one which is
defined according to $\tilde{\Gamma}_m$, where $\Gamma_n \subseteq \tilde{\Gamma}_m$,
$n \leq m$, i.e. $\tilde{\Gamma}_m$ contains $\Gamma_n$ as a subgraph. 
One obtains 
\begin{equation}
\Psi_{\Gamma_n,f_n}(A) = \Psi_{\tilde{\Gamma}_m,\tilde{f}_m}(A) \;,
\end{equation}
simply by requiring $\tilde{f}_m$ to be independent of the $(m-n)$ group 
elements $U_i$ which belong to links in $\tilde{\Gamma}_m$ but not to $\Gamma_n$.
In other words, any two cylindrical functions can always be viewed as being 
defined on the same graph which is just constructed as the union of the original 
ones.
Given this property, it is now straightforward to define a scalar product for 
any two cylindrical functions by
\begin{equation}
\label{scalar_product}
  \left< \Psi_{\Gamma_n,f_n} |\, \Psi_{\Gamma_n,g_n} \right> := \int_{[SU(2)]^n}
  dU_1 \ldots dU_n \: \overline{f_n(U_1, \ldots, U_n)} \: g_n(U_1, \ldots, U_n) \;.
\end{equation}
Here $dU_1 \ldots dU_n$ is the induced Haar measure on $[SU(2)]^n$.
This scalar product extends by linearity and continuity to a well-defined scalar product 
on $L$. To simplify the notation, we will drop the index $n$ from now on.

The unconstrained Hilbert space $\mathcal{H}$ of quantum states 
is obtained by completing the space of all finite linear combinations 
of cylindrical functions (for which the scalar product is also defined) in 
the norm induced by the quadratic form (\ref{scalar_product}) on
a cylindrical function as $\|\Psi_{\Gamma,f}\| \, = 
\left< \Psi_{\Gamma,f} |\, \Psi_{\Gamma,f} \right>^{1/2}$.
This (non-separable) Hilbert space $\mathcal{H}$\footnote{In the literature, the 
Hilbert completion of $L$ in the scalar product (\ref{scalar_product}) is often 
referred to as the auxiliary Hilbert space $\mathcal{H}_{aux}$.}
on which loop quantum gravity is defined, has the properties 
we required in the beginning of this section, namely it carries a unitary 
representation of local $SU(2)$, and a unitary representation of $Diff (M)$.
These unitary representations on $\mathcal{H}$ are naturally realized on 
the quantum states $\Psi(A) \in \mathcal{H}$ by transformations of 
their arguments.

Under (smooth) local $SU(2)$ gauge transformations 
$\lambda : M \to SU(2)$ the connection $A$ transforms
inhomogeneously like a gauge potential, i.e.
\begin{equation}
\label{connection_trafo}
A\to A_{\lambda} = \lambda^{-1} A \lambda + \lambda^{-1} d\lambda \;.
\end{equation}
This induces a natural representation of local gauge 
transformations $\Psi(A) \to \Psi(A_{\lambda})$ on $\mathcal{H}$.
Similarly, if one considers spatial diffeomorphisms $\phi: M \rightarrow M$, one 
finds that the connection transforms as a 1-form, 
\begin{equation}
A \to \phi^{-1} A \;.
\end{equation}
Hence $\mathcal{H}$ carries a natural representation of $Diff(M)$ via
$\Psi(A) \to \Psi(\phi^{-1} A)$.

The fact that $\mathcal{H}$ carries \emph{unitary} representations stems from the
invariance of the scalar product (\ref{scalar_product}) under local $SU(2)$ 
transformations and spatial diffeomorphisms which can be seen as follows.
Despite the inhomogeneous transformation rule (\ref{connection_trafo}) of 
the connection under gauge transformations, the holonomy turns out to 
transform homogeneously like 
\begin{equation}
U[A,\gamma] \to U[A_{\lambda},\gamma] = 
\lambda^{-1}(x_i)\, U[A,\gamma]\, \lambda(x_f) \;,
\end{equation}
where $x_i, x_f \in M$ are the initial and final points of the curve $\gamma$,
respectively. Now, since a cylindrical function depends on the connection $A$ only 
via the holonomy, it transforms as 
\begin{equation}
f(U_1, \ldots, U_n) \to f_{\lambda}(U_1, \ldots, U_n) = f(\lambda^{-1} U_1 \lambda, 
\ldots, \lambda^{-1} U_n \lambda) \;.
\end{equation}
Writing this in terms of quantum states, we obtain 
\begin{equation}
\Psi_{\Gamma, f}(A_{\lambda}) = \Psi_{\Gamma, f_{\lambda}}(A) \;.
\end{equation}
This shows immediately the invariance of (\ref{scalar_product}) under gauge
transformations since the Haar measure is by definition $[SU(2)]^n$ invariant.

Considering diffeomorphisms, the transformation of the holonomy is induced as 
\begin{equation}
U[A,\gamma] \to U[\phi^{-1} A,\gamma] = U[A, \phi \cdot \gamma] \;,
\end{equation}
which means nothing but a shift of the curve $\gamma$ along which the holonomy is 
defined. Hence a diffeomorphism transforms a quantum 
state $\Psi(A)$ into one which is based on a shifted graph. Since the right hand 
side of (\ref{scalar_product}) does not depent explicitely on the graph, the
diffeomorphism invariance of the inner product is obvious.

There are several mathematical developements connected with the 
construction given above. They involve projective families 
and projective limits, generalized connections, representation 
theory of $C^*$-algebras, measure theoretical techniques, and 
others.  These developments, however, are not needed 
for the following, and for understanding the basic physical 
results of loop quantum gravity.  For details and further references on 
these developements, we refer to \cite{Lewand_Karpacz99} and \cite{Ashtekar93}.
 
\subsection{A Basis in the Hilbert Space}
We now construct an orthonormal basis in the Hilbert space 
$\mathcal{H}$. We begin by defining a \emph{spin network}, which 
is an extension of the notion of graph, namely a colored graph.  
Consider a graph $\Gamma$ with $n$ links $\gamma_i$, $i=1, 
\ldots, n$, embedded in the 3-manifold $M$.  To each link 
$\gamma_i$ we assign a non-trivial irreducibel representation 
of $SU(2)$ which is labeled by its spin $j_i$ or equivalentely by 
$2 j_i$, an integer which is called the \emph{color} of the link.  
The Hilbert space on which this irreducible spin-$j_i$ 
representation is defined is denoted as $\mathcal{H}_{j_i}$.

Next, consider a particular node $p$, say a $k$-valent one.  
There are $k$ links $\gamma_1, \ldots, \gamma_k$ that meet at 
this node. They are colored by $j_{1}, \ldots, j_{k}$.  Let
\mbox{$\mathcal{H}_{j_1}, \ldots, \mathcal{H}_{j_k}$} be the Hilbert 
spaces of the representations associated to the $k$ links.  
Consider the tensor product of these spaces
\begin{equation}
\label{Hp}
\mathcal{H}_{p} = \mathcal{H}_{j_1} \otimes \ldots \otimes 
\mathcal{H}_{j_k} \; ,
\end{equation}
and fix, once and for all, an orthonormal basis in $\mathcal{H}_{p}$.  
This choice of an element $N_{p}$ of the basis is called a 
\emph{coloring of the node $p$}.  

A (non-gauge invariant) \emph{spin network} $S$ is then defined 
as a colored embedded graph, namely as a graph embedded in 
space in which links as well as nodes are colored.  More precisely, it 
is an embedded graph plus the assignement of a spin $j_{i}$ to 
each link $\gamma_{i}$ and the assigmenet of an (orthonormal) basis element 
$N_{p}$ to each node $p$. A spin network is thus a triple 
$S=\left( \Gamma, \vec{j}, \vec{N} \right)$.  The vector 
notations $\vec{j}$ and $\vec{N}$ are abbreviations for $\vec{j} 
= \{j_i \}$, $i=1, \dots, n$, the collection of all irreducible 
$SU(2)$ representations associated to the $n$ links in $\Gamma$, 
and $\vec{N} = \{N_p \}$ stands for the basis elements attached 
to the nodes.

Now we are able to define a \emph{spin network state} $\Psi_S(A)$ 
as a cylindrical function $f_S$ associated to the spin network 
$S$ whose graph is $\Gamma$, as
\begin{equation}
\label{statedef}
\Psi_S(A) = \Psi_{\Gamma, f_S} (A) = 
        f_S(U[A,\gamma_1], \ldots, U[A,\gamma_n]) \;.
\end{equation}
The cylindrical function $f_S$ is constructed by taking the 
holonomy along each link of the graph in that irreducible 
representation of $SU(2)$ which is associated to the link. 
The holonomy matrices are contracted with the vector $N_{p}$ at each 
node $p$ where the links meet, giving 
\begin{equation}
\label{cyl}
\Psi_S(A) =  f_S(U_1, U_2, \ldots, U_n) = \bigotimes_{\mbox{\scriptsize links } 
	i \in \Gamma} R^{j_i}(U_i) \, \cdot \bigotimes_{\mbox{\scriptsize nodes } 
	p \in \Gamma} N_p \; .
\end{equation} 
Here $R^{j_i}(U_i)$ is the representation matrix of the holonomy or group element
$U_i$, respectively, in the spin-$j_i$ irreducible representation of $SU(2)$ 
associated to a link $\gamma_i$, $i=1,\dots,n$.  
It is considered as an element of $\mathcal{H}^{j_i} \otimes \mathcal{H}^{j_i}$.
Since $N_p$ is in the tensor product (\ref{Hp}) 
of the Hilbert spaces associated to the links that meet at a 
node, i.e. it can be seen as a tensor with one index in each 
of these spaces, the identification
\begin{equation}
  \bigotimes_{\mbox{\scriptsize nodes } p \in \Gamma} \mathcal{H}_p
  = \bigotimes_{\mbox{\scriptsize links } i \in \Gamma} 
  \left(\mathcal{H}^{j_i} \otimes \mathcal{H}^{j_i} \right)
\end{equation}
is implied.
Hence the dot $\cdot$ in (\ref{cyl}) indicates the scalar product in
$\bigotimes_{\mbox{\scriptsize links } i \in \Gamma} 
\left(\mathcal{H}^{j_i} \otimes \mathcal{H}^{j_i} \right)$,
and one recognizes the exact matching of the indices.

By varying the graph, the colors of the links, and the basis 
elements (i.e. colors) at the nodes, we obtain a family of states, which 
turn out to be normalized.  At last, using the well-known Peter--Weyl 
theorem, it can easily be shown that any two distinct states 
$\Psi_S$ are orthonormal in the scalar product (\ref{scalar_product}),
\begin{eqnarray}
\langle \Psi_S | \Psi_{S'}\rangle &=& \delta_{SS'} \\ 
 &=& \delta_{\Gamma, \Gamma'} \delta_{\vec{j}, \vec{j'}} 
\delta_{\vec{N}, \vec{N'}} \; ,
\end{eqnarray}
and that if $\Psi$ is orthogonal to every spin network state, then 
$\Psi=0$.  Therefore the spin network states form a complete 
orthonormal (non-countable) basis in the kinematical Hilbert space $\mathcal{H}$.

Just as in conventional quantum mechanics, one can distinguish an algebraic 
as well as a differential version of quantum gravity.
The best example from quantum mechanics is probably the harmonic oscillator.
There one can consider the Hilbert space of square integrable functions 
on the real line, and express the momentum and the Hamiltonian as differential 
operators. Solving the Schr\"odinger differential equation explicitely gives the 
eigenstates of the Hamiltonian which may be denoted as 
$\Psi_n(x) = \langle x | n \rangle$. As the reader certainly knows, the theory can be 
expressed entirely in algebraic form in terms of the states $| n \rangle$. In that 
case, all elementary operators are purely algebraic in nature.
A similar scheme applies also to quantum gravity. There one can define the theory by
working exlusively in the spin network (or loop) basis, without ever mentioning 
functionals of the connection. This (algebraic) representation of the theory is 
called \emph{loop representation}. On the other hand, using wave functionals 
$\Psi(A)$ defines the differential version of the theory, known as 
\emph{connection representation}. The relation between both formalism may be
expressed by $\Psi_S(A) = \langle A | S \rangle$ which defines 
a unitary mapping. The abstractly defined spin network
basis $|S \rangle$ will turn up in a slightly different (namely gauge invariant) 
context in sect.~\ref{sect_dynamics} below.
The (unitary) equivalence of both versions of quantum gravity was proven by 
De Pietri in \cite{DePietri97}.

\subsection{The $SU(2)$ Gauge Constraint}
\label{su2}
The physical quantum state space $\mathcal{H}_{phys}$ is obtained 
by imposing the quantum constraint equations on the Hilbert space 
$\mathcal{H}$. We want to impose the quantum constraints one 
after another as it is shown in Fig.~\ref{construction_of_H}.

\vspace*{0.5cm}
\begin{figure}[h]
\begin{tabbing}
\hspace*{3.5cm}\=$\mathcal{H}$ \quad \=$\stackrel{SU(2)}{\longrightarrow}$ 
\quad \=$\mathcal{H}_0$ 
\quad \=$\stackrel{\mbox{\scriptsize\emph{Diff(M)}}}{\longrightarrow}$ 
\quad \=$\mathcal{H}_{diff}$ \quad \=$\stackrel{\widehat{H}}{\longrightarrow}$ 
\quad \=$\mathcal{H}_{phys}$ \\
\\
\>\hspace*{1mm}$\downarrow$ \>\>\hspace*{2mm}$\downarrow$ 
\>\>\hspace*{3mm}$\downarrow$ \>\>\hspace*{3mm}$\downarrow$ \\
\\
\>\hspace*{-1mm}$\Psi(A)$ \>\hspace*{1mm}$\longrightarrow$ \>$|S\rangle$ 
\>\hspace*{2mm}$\longrightarrow$ \>\hspace*{2mm}$|s\rangle$
\>$\longrightarrow$ \> \hspace*{2mm} {\large\bfseries ?} 
\end{tabbing}
\caption{A step by step construction of the physical Hilbert space.}
\label{construction_of_H}
\end{figure}

In this diagram the first line refers to the imposition of the 
quantum constraints yielding the appropriate invariant Hilbert spaces, 
while the second line shows the corresponding basis. 
The question mark stands for the fact that the explicit construction of the 
states in the physical Hilbert space is not yet understood.  
This is not surprising, since 
having the complete set of these states explicitely would amount 
to having {\em solved\/} the theory completely, a much 
stronger result than what we are looking for.

We begin here the process of solving the constraints by first considering
the $SU(2)$ gauge constraint.  According to what we described in 
sect.~\ref{subsect_cylfct} the transformation properties of the quantum states 
$\Psi(A)$ under local $SU(2)$ gauge transformations $\lambda(x)$ 
are easy to work out.  In fact, a moment 
of reflection shows that a gauge transformation acts on 
a spin network state simply by $SU(2)$ transforming the coloring 
of the nodes $N_{p}$. More precisely, the spaces 
$\mathcal{H}_{p}$, which carry a representation of $SU(2)$, are 
transformed by the $SU(2)$ element $\lambda(x_{p})$, where 
$x_{p}$ is the point of the manifold in which the node $p$ lies.

It is then easy to find the complete set of gauge invariant 
states. The Hilbert space $\mathcal{H}_{p}$, being a tensor 
product of irreducible representations, can be decomposed into its 
irreducible parts,
\begin{equation}
\label{decomposition}
\mathcal{H}_{j_1} \otimes \ldots \otimes \mathcal{H}_{j_k} = \bigoplus_J 
\left(\sH_J \right)^{k_J} \; .
\end{equation}
Here $k_J$ denotes the multiplicity of the spin-$J$ irreducible representation.
Among all sub\-spaces of this decomposition we are interested in the 
SU(2) gauge invariant one (the singlet), i.e. the $J=0$ subspace,
denoted as $\left( \sH_0 \right)^{k_0}$  (not to be confused with the
Hilbert space $\mathcal{H}_0$ in Fig.~\ref{construction_of_H}). 
We pick an arbitrary basis in this subspace,
and assign one basis element $N_p$ to the node $p$.
A spin network in which the 
coloring of the nodes is given by such invariant tensors $N_{p}$ 
is called a \emph{gauge invariant spin network} (often, the expression 
spin network is used for the gauge invariant ones). The corresponding 
spin network states constructed in terms of gauge invariant spin 
networks solve the gauge constraint and form a complete 
orthonormal basis in $\mathcal{H}_{0}$, the $SU(2)$ gauge invariant Hilbert space. 

The quantities $N_p$ are called \emph{intertwiners}. They are 
invariant tensors with indices in different irreducible $SU(2)$ representations
which provide the possibility to couple representations of 
$SU(2)$. In other words, they map the incoming irreducible representations 
at a node to the outgoing ones\footnote{Denoting indices as \lq \lq 
incoming'' and \lq \lq outgoing'' is just a convenient labelling.
Actually one may wonder why we don't really care about the 
orientation of the links in the graph.  This happens just because 
it can be neglected in the case where $SU(2)$ acts as gauge group, 
being a consequence of the following.  In general, an inversion 
of the orientation of a link $\gamma$ with associated irreducible 
representation $j$ would lead to a change of this representation 
to its conjugate $j^*$. But since for $SU(2)$ $j$ and $j^*$ are 
unitary equivalent, considerations concerning the orientation are
simplified, and we don't really have to worry about it.}. 
Thus they are given by standard Clebsch--Gordon theory.

\begin{figure} 
\centerline{\includegraphics[width=10cm,angle=0]{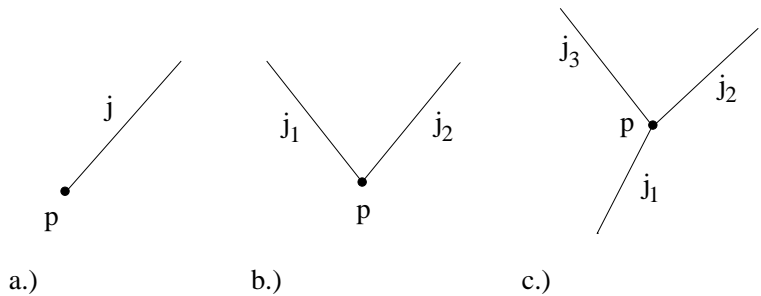}}
\caption[]{Examples of $n=1,2,3$-valent nodes.}
\label{intertwiners}
\end{figure}

To clarify the mathematics of the intertwiners, consider some 
examples.  In the case of a 1-valent node as shown in 
Fig.~\ref{intertwiners}a, there is no invarinat tensor, hence 
the dimensionality of the corresponding Hilbert space 
$\sH_0$ is zero.  Consider on the other hand 
Fig.~\ref{intertwiners}b with a 2-valent node $p$, there exists a 
single intertwiner only if the colors of the links are equal, 
which is
\begin{equation}
(N_p)_{j_1 j_2} = \delta_{j_1 j_2} \; .
\end{equation}
The last and most interesting example is Fig.~\ref{intertwiners}c with a
trivalent node, which
corresponds to the coupling of three spins, well-known from the quantum theory of 
angular momentum. As long as the representations associated to the links 
satisfy the Clebsch--Gordan condition $|j_2 - j_3| \leq j_1 \leq j_2 + j_3$,
once $j_2$ and $j_3$ are fixed (analogously for any other pair of $j$'s), 
a unique intertwiner exists because there is only one way of combining three
irreducible representations in order to obtain a singlet.
The invariant tensor is then given by nothing but the familiar 
Wigner $3j$-coefficient, which is (apart from normalization)
\begin{equation}
(N_p)_{n_1 n_2 n_3} = 
        \left( \begin{array}{ccc}
                j_1 & j_2 & j_3 \\
                n_1 & n_2 & n_3 \\
        \end{array} \right) \; .
\end{equation}
If the Clebsch-Gordan condition is not satisfied,
the dimension of $\sH_0$ is zero again.

\begin{figure}
\centerline{\includegraphics[width=7cm]{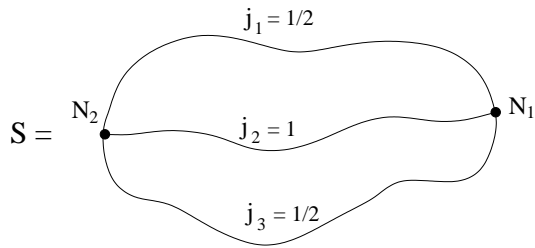}}
\caption[]{A simple spin network with two trivalent nodes.}
\label{spin_net1}
\end{figure}

Let's now consider a simple example of a spin network state.
We take the spin network in Fig.~\ref{spin_net1} corresponding to a
graph with two trivalent nodes and three links joining them.
Let two of the links carry (fundamental) spin-1/2 representations, while 
the third link has a spin-1 representation attached to it.

The elements $N_1$ and $N_2$ of an appropriate basis of invariant 
tensors are assigned to the nodes.
The corresponding spin network state then reads explicitely

\begin{eqnarray}
\Psi_S(A) &=& R^\frac{1}{2}[U_1]_A{}^B \, 
        R^1[U_2]_i{}^j \,
        R^\frac{1}{2}[U_3]_C{}^D \, (N_1)^{A i C} \, (N_2)_{B j D} \nonumber \\
	\label{example}
        &=& (U_1)_A{}^B \, R^1[U_2]_i{}^j \, (U_3)_C{}^D \,  
        \left( \begin{array}{ccc}
                 \scriptstyle{\frac{1}{2}} & \scriptstyle{1} 
                 & \scriptstyle{\frac{1}{2}} \\
                 \scriptstyle{A} & \scriptstyle{i} & \scriptstyle{C} \\
        \end{array} \right) \, 
        \left( \begin{array}{ccc}
                 \scriptstyle{\frac{1}{2}} & \scriptstyle{1} 
                 & \scriptstyle{\frac{1}{2}} \\
                 \scriptstyle{B} & \scriptstyle{j} & \scriptstyle{D} \\
        \end{array} \right) \;.
\end{eqnarray}
Here $i,j = 1,2,3$ denote vector and $A,B, \ldots = 1,2$ spinor indices, respectively.
The holonomy is abbreviated as $U_k \equiv U[A,\gamma_k]$.

Finally, we mention that each spin network state can be
decomposed into a finite linear combination of products of loop states.
This decomposition can be done in general by using the 
following rule, which follows from well-known properties of 
$SU(2)$ representation theory.  Replace each link of the graph 
with associated spin $j$ with $2j$ parallel strands.  Antisymmetrize these 
strands along each link (obtaining a formal linear combination 
of drawings).  The intertwiners at the nodes can be represented 
as collections of segments joining the strands of different 
links.  By joining these segments with the strands one obtains a 
linear combination of multiloops.  The spin network states can 
then be expanded in the corresponding loop states. For details of 
this construction, see \cite{RovelliSmolin95b,DePietriRovelli96}.

\begin{figure}
\centerline{\includegraphics[width=12cm]{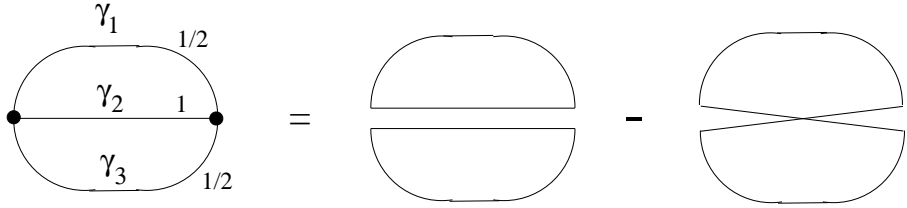}}
\caption[]{The decomposition of the spin network state (\ref{example}) into loop states.}
\label{spinnet_decomp}
\end{figure}

Applying this rule to the above example (\ref{example}), we obtain the
following. Writing out the explicit expression for 
the spin-1 representation in terms of spin-1/2 representations 
(which we will not do here), and using the explicit form of the 
Clebsch--Gordan coefficient, it is not hard to see that 
\begin{equation}
\Psi_S(A) = \Psi_{\alpha} - \Psi_{\beta} \;, 
\end{equation}
where $\beta$ is the loop obtained by joining the four segments 
$\gamma_{1}$, $\gamma_{2}$, $\gamma_{3}$, $\gamma_{2}$, and 
$\alpha$ is the double loop $\{\alpha_{1},\alpha_{2}\}$. Here 
$\alpha_{1}$ is obtained by joining $\gamma_{1}$ and $\gamma_{2}$, while 
$\alpha_{2}$ is obtained by joining $\gamma_{2}$ and $\gamma_{3}$.  
For a graphical illustration, see Fig.~\ref{spinnet_decomp}. 

\subsection{Operators on $\mathcal{H}$}
\label{ops_on_H}
We now have a gauge invariant kinematical Hilbert space of quantum
gravity including an orthonormal basis of spin network states at our disposal.
Below, we want to construct self-adjoint operators corresponding to the classical fields.

In this subsection we will straightforwardly construct well-defined 
gauge invariant operators and think about their physical interpretation 
in the next section.
We proceed as in usual quantum mechanics by constructing multiplicative
and derivative operators, corresponding to \lq \lq position'' and \lq \lq momentum'',
respectively. See also \cite{DePietriRovelli96} and \cite{RovelliUpadhya98}.

The simplest operator is given by the holonomy itself. 
Given a curve $\gamma$, take the holonomy of the connection along $\gamma$ to define
the multiplicative operator
\begin{equation}
 \hat{U}(\gamma) = U[A, \gamma] \;.
\end{equation}
More precisely, $\hat{U}(\gamma)$ defines a matrix-valued operator.
This operator is not $SU(2)$ gauge invariant.
In order to obtain gauge invariance, 
we simply consider the (negative\footnote{See footnote \ref{minus} 
in sect. \ref{sect_Holonomies}.}) trace of 
the ho\-lo\-no\-my along a loop $\alpha$, 
resulting in the operator
\begin{equation}
 \hat{T}[\alpha] = - \mbox{Tr}\,U[\alpha] \;.
\end{equation}
This definition provides a well-defined, gauge invariant and multiplicative 
operator\footnote{It is really the choice of these so-called 
\emph{Wilson loops} which is the characteristic
feature of loop quantum gravity. Indeed, the loop approach can be built on 
Wilson loops and appropriate momentum operators (the so-called loop variables) which 
form a closed algebra and were thus used as the starting point for canonical 
quantization.} acting on a state functional as
\begin{equation}
\label{Wilsonloop_op}
 \hat{T}[\alpha] \, \Psi_S(A) = - \mbox{Tr}\left( \mathcal{P}\, \exp\int_{\alpha} 
	A\right) \, \Psi_S(A) = - \mbox{Tr}\, U[A, \alpha] \, \Psi_S(A) \;. 
\end{equation}
Hence the definition of multiplicative operators doesn't seem to cause 
any problems.

The construction of a gauge invariant derivative operator turns out to be more subtle.
The configuration variable in our approach is the connection $A(x)$, thus
the conjugate momentum operator would be some functional derivative with respect
to it. The same statement is obtained by first considering
the Poisson algebra (\ref{Poissonalgebra}) and proceeding as 
usual in quantum field theories, i.e. by formally replacing $E^a_i$ with the 
functional derivative \linebreak
(we neglect here the Immirzi parameter $\beta$),
\begin{equation}
\label{Eai}
   E^a_i(x) \; \longrightarrow \; -\, i \hbar G \, \frac{\delta}{\delta A^i_a (x)} \;.
\end{equation}
This object is an operator-valued distribution rather than a 
genuine operator, so it has to be integrated against test functions, or, in other
words, it has to be suitably smeared in order to be well-defined.
Thus, to transform (\ref{Eai}) into a genuine operator and 
regularize expressions involving it, an appropriate smearing over a 
surface $\Sigma$ has to be performed.
The use of 2-dimensional surfaces rather than 3-dimensional ones (as one might have 
guessed first) can roughly be seen as follows.
The functional derivative (\ref{Eai}) 
with respect to the connection 1-form $A(x)$ is a vector density  
of weight one, or equivalentely a 2-form. Contracting the vector density $E^a$
with the Levi--Civita density gives the dual of the triad, which is a 2-form
$E=\epsilon_{abc} E^a dx^b dx^c$. Hence they may be identified.
Since 2-forms are naturally integrated against 2-surfaces, 
a geometrical, i.e. coordinate or background independent regularization scheme,
respectively, is naturally suggested.
And indeed, it turned out to be the most convenient way of handling the problem!
In fact, smearing (\ref{Eai}) as described above, will give us a well-defined
operator which is \emph{coordinate invariant} and \emph{finite}.

We start by considering a surface, that is a 2-dimensional manifold $\Sigma$ 
embedded in $M$.
We use local coordinates $x^a$, $a=1,2,3$, on $M$ and let 
$\vec{\sigma} = (\sigma^1, \sigma^2)$ be coordinates on the surface $\Sigma$.
Thus the embedding is given by
\begin{equation}
\Sigma:\; (\sigma^1, \sigma^2) \mapsto x^a(\sigma^1, \sigma^2) \;.
\end{equation}
We define an operator (using $G = \hbar = 1$)
\begin{equation}
  \label{E_operator}
  \hat{E}^i(\Sigma):= - i \int_\Sigma d\sigma^1 d\sigma^2 \, n_a(\vec{\sigma})\, 
        \frac{\delta}{\delta A^i_a\big(x(\vec{\sigma})\big)} \;,
\end{equation}
where
\begin{equation}
  n_a(\vec{\sigma}) = \epsilon_{abc} \, \frac{\partial x^b(\vec{\sigma})}{\partial \sigma^1}
        \frac{\partial x^c(\vec{\sigma})}{\partial \sigma^2}        
\end{equation}
is the normal 1-form on $\Sigma$ and $\epsilon_{abc}$ is the Levi-Civita tensor of 
density weight $(-1)$.

The next step is to compute the action of this operator
on holonomies $U[A,\gamma]$, which are the basic building blocks
of the gauge invariant state functionals, i.e. the spin network states. The 
coordinates of the curve $\gamma$ in $M$, which is parametrized by $s$, 
will be denoted in the following as $x^a(s) \equiv \gamma^a(s)$. 

\begin{figure}[t]
\centerline{\includegraphics[width=7cm,angle=0]{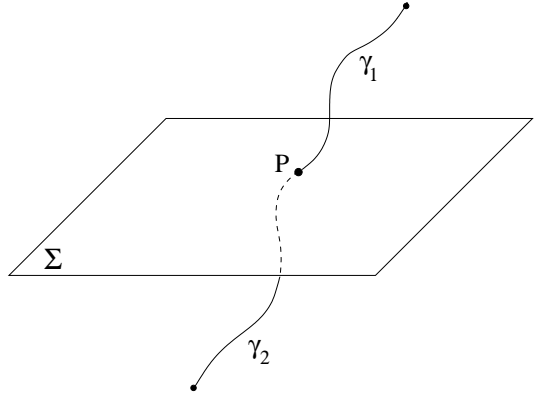}}
\caption[]{A curve that intersects the surface in an individual point.}
\label{intersection1}
\end{figure}

We begin with the functional derivative of holo\-no\-mies. A 
detailed derivation of the relevant formulas can be obtained 
using the first variation of the defining differential equation 
(\ref{paralleltransp}) of the holonomy with respect to the 
connection, see \cite{LewandNewmanRovelli93} for further details. 
Consider the surface $\Sigma$ and a curve $\gamma$ along which the holonomy 
is constructed in the simplest case where they have only one 
individual point of intersection $P$. Furthermore $P$ is not supposed to 
lie at the endpoints of $\gamma$, as shown in 
Fig.~\ref{intersection1}.  

For later convenience the curve is devided into two parts, $\gamma = \gamma_1 \cup \gamma_2$,
one lying \lq \lq above'', the other \lq \lq below'' the surface.
We get
\begin{eqnarray} 
\frac{\delta}{\delta A^i_a (\vec{x}(\vec{\sigma}))} \, U[A,\gamma] &=& 
\frac{\delta}{\delta A^i_a (\vec{x}(\vec{\sigma}))} \, \left( \mathcal{P} 
        \exp \int_{\gamma} ds \, \dot{x}^a \, A_a^i(\vec{x}(s)) \, \tau_i \right) 
        \nonumber \\ 
        \label{holonomyderiv}
        &=& \int_{\gamma} ds \, \frac{\partial x^a}{\partial s} \,
        \delta^3 \big( \vec{x}(\vec{\sigma}), \vec{x}(s) \big) \,
        U[A,\gamma_1]\, \tau_i \, U[A,\gamma_2] \;.
\end{eqnarray}
Here, $U[A,\gamma_1]$ and $U[A,\gamma_2]$ are the parallel propagators along those 
segments of $\gamma$ which \lq \lq start'' or \lq \lq end'', respectively, on 
$P = \Sigma\, \cap \,\gamma \not= \emptyset$, see Fig.~\ref{intersection1}. 
In order to avoid confusion, recall that $\vec{x}(\vec{\sigma})$ are the 
coordinates of the surface $\Sigma$ embedded in the 3-manifold $M$, 
while $\vec{x}(s)$ are the coordinates of $\gamma=\gamma_1 \cup \gamma_2$, 
just as defined in sect.~\ref{sect_Holonomies}.

We are now prepared to care about the action of the operator $\hat{E}^i(\Sigma)$ on
$U[A,\gamma]$. Using (\ref{holonomyderiv}), the calculation can immediately be performed,
giving
\begin{eqnarray}
  \lefteqn{\hat{E}^i(\Sigma) \, U[A,\gamma]} \hspace{0.5cm} \nonumber \\
  &=& -i \int_{\Sigma} d\sigma^1 d\sigma^2 \,\epsilon_{abc} \,
  \frac{\partial x^a(\vec{\sigma})}{\partial \sigma^1}
  \frac{\partial x^b(\vec{\sigma})}{\partial \sigma^2}
  \, \frac{\delta}{\delta A^i_c\big(\vec{x}(\vec{\sigma})\big)} \:
  U[A,\gamma] \nonumber \\
  \label{EonU}
  &=& -i \int_{\Sigma} \int_{\gamma} d\sigma^1 d\sigma^2 ds \:\epsilon_{abc}\,
  \frac{\partial x^a}{\partial \sigma^1}\frac{\partial x^b}{\partial \sigma^2}
  \frac{\partial x^c}{\partial s} \, 
  \delta^3 \big( \vec{x}(\vec{\sigma}), \vec{x}(s) \big) \times \nonumber \\
  && \hspace{5cm} \times \: U[A,\gamma_1]\, \tau^i \, U[A,\gamma_2] \;.
\end{eqnarray}
A closer look at this result reveals a great simplification of the last integral since
one notices the appearance of the Jacobian $J$ for the 
transformation of the coordinates \linebreak
\mbox{$(\sigma^1, \sigma^2, s) \rightarrow (x^1, x^2, x^3)$}, namely
\begin{equation}
  \label{jacobian}
  J \equiv \frac{\partial\,(\sigma^1, \sigma^2, s)}{\partial\,(x^1, x^2, x^3)}
  = \epsilon_{abc}\, \frac{\partial x^a}{\partial \sigma^1}
  \frac{\partial x^b}{\partial \sigma^2}\frac{\partial x^c}{\partial s} \;.
\end{equation}
In our case, we may assume that the Jacobian is non-vanishing, since we have required 
that only a single, non-degenerate point of intersection 
of $\Sigma$ and $\gamma$ exists. The Jacobian (\ref{jacobian})
and the integral (\ref{EonU}) would vanish, if the tangent vectors 
given by the partial derivatives in (\ref{jacobian}), would be coplanar, i.e. if
a tangent $\partial x^{a,b}(\vec{\sigma}) / \partial \sigma^{1,2}$ to the surface
would be parallel to the tangent $\partial x^c(s) / \partial s$ of the curve.
This happens, for instance, if the curve lies entirely in $\Sigma$. Then 
there would be of course no individual \emph{point} of intersection.
We will consider the various cases a little closer at the end of this section.

But let's come back to the actual topic---the simplification 
of (\ref{EonU}). Carrying out the described
coordinate transformation will put us in the position
to integrate out the \mbox{3-dimensional} \mbox{$\delta$-distribution}. 
We get for the case of a single intersection the 
interesting result\footnote{The integer-valued integral is indeed
an analytic (coordinate independent) expression for the intersection number 
of the surface $\Sigma$ and the curve $\gamma$.
It is zero in the case of no intersection at all.}
\begin{equation}
  \int_{\Sigma} \int_{\gamma} d\sigma^1 d\sigma^2 ds \:\epsilon_{abc}\,
  \frac{\partial x^a(\vec{\sigma})}{\partial \sigma^1}
  \frac{\partial x^b(\vec{\sigma})}{\partial \sigma^2}
  \frac{\partial x^c(s)}{\partial s} \,
  \delta^3 \big( \vec{x}(\vec{\sigma}), \vec{x}(s) \big) \, = \pm 1 \;,
\end{equation}
where the sign depends on the relative orientation of the surface to the 
curve (this sign will soon become irrelevant).
Hence we obtain the simple result
\begin{equation}
  \hat{E}^i(\Sigma) \, U[A,\gamma] = \pm i\, U[A,\gamma_1]\, \tau^i \, U[A,\gamma_2] \;.
\end{equation}
So we see finally that the action of the operator $\hat{E}^i(\Sigma)$ on holonomies
consists of just inserting the matrix $(\pm i \, \tau^i)$ at the point of intersection.
Taking advantage of this result, the generalization to the case of more than 
one single point of intersection is trivial---it is just the sum of all such insertions.

Putting all this together, and using $P$ to denote different separate 
points of intersection, we have:
\begin{equation}
\label{EonU_2}
  \hat{E}^i(\Sigma) \, U[A,\gamma] = \left\{
        \begin{array}{c@{\qquad}l}
                0 & \mbox{if\quad} \Sigma \cap \gamma = \emptyset \\
                \vspace{0mm} & \\
                 \sum\limits_{P} \pm i\, U[A,\gamma^P_1]\, 
                        \tau^i \, U[A,\gamma^P_2] & \mbox{if\quad} P \in \Sigma \cap \gamma \;.
        \end{array} \right.    
\end{equation}
A further generalization of (\ref{EonU_2}) is needed in view of spin networks, 
where arbitrary irreducible spin-$j$ representations are associated to links 
and the accompanying holonomies, denoted as $R^j \big( U[A,\gamma]\big)$.   
We obtain easily (again for just one single point of intersection, 
which may be extended analogously to (\ref{EonU_2}))
\begin{equation}
\label{EonU_3}
  \hat{E}^i(\Sigma) \, R^j \big( U[A,\gamma] \big) = \pm i \, R^j \big( U[A,\gamma_1]\big) 
        \, {}^{\scriptscriptstyle (j)}\tau^i \, R^j \big( U[A,\gamma_2]\big) \;.
\end{equation}
Here ${}^{\scriptscriptstyle (j)}\tau^i$ is the corresponding $SU(2)$ 
generator in the spin-$j$ representation.

We now have a well-defined operator at our disposal.
One may again wonder why this smearing scheme gives a 
well-defined operator, since we have used only a 2-dimensional smearing 
over a surface $\Sigma$ instead of a 3-dimensional one over $M$,
as one might have expected.
But the clue to this is that the state functionals have support on one dimension, or in
other words, they contain just 1-dimensional excitations.

The action of $\hat{E}^i (\Sigma)$ on a spin network state 
$\Psi_S(A)$ follows immediately from the above 
considerations.  We take a gauge invariant spin network $S$ 
which intersects the surface $\Sigma$ at a single point.  The 
holomony along the crossing link $\gamma$ being in the 
spin-$j$ representation of $SU(2)$.  
Splitting the spin network state $\Psi_S(A)$ into a part which consists of
this holonomy $R^j \big( U[A,\gamma] \big)$ along $\gamma$,
and the \lq \lq rest'' of the state\footnote{To see how this is possible 
recall the definitions (\ref{statedef}) and (\ref{cyl}) of a spin network 
state as a cylindrical function.}, which is denoted as $\Psi_{S-\gamma}(A)$, 
we get
\begin{equation}
  \Psi_S(A) = \Psi^{mn}_{S-\gamma}(A)\: R^j \big( U[A,\gamma] \big)_{mn} \;.
\end{equation}
Here we have used the index notation with $m$ and $n$ being indices in the 
Hilbert space that is attached to $\gamma$.
Obviously $\Psi_{S-\gamma}(A)$ is not gauge invariant any more.
Using (\ref{EonU_3}) we get immediately
\begin{equation}
\label{EonPsi}
  \hat{E}^i (\Sigma) \, \Psi_S(A) = \pm i   
  \Big[ R^j \big( U[A,\gamma_1]\big) \, {}^{\scriptscriptstyle (j)}\tau^i  \, 
  R^j \big( U[A,\gamma_2]\big) \Big]_{mn} \, \Psi^{mn}_{S-\gamma}(A) \;.  
\end{equation}
Eventually, we see that $\hat{E}^i (\Sigma)$ spoils gauge invariance, since the resulting
functional is not an element of $\mathcal{H}_0$ any more.
The construction of a gauge invariant derivative operator will be shown 
in the next section.

\subsubsection{An $SU(2)$ Gauge Invariant Operator.}
Gauge invariance is spoiled in (\ref{EonPsi}) by the 
insertion of a matrix $\tau_{i}$ (which is gauge 
covariant, but not gauge invariant) at the point of 
intersection.  We can try to construct a gauge invariant operator 
simply by squaring this matrix, namely by defining 
\begin{equation}
  \hat{E}^{\,2}(\Sigma) := \hat{E}^i (\Sigma)\,\hat{E}^i (\Sigma) \;,
\end{equation}
where summation over $i = 1,\dots, 3$ is assumed.
Let us compute the action of this operator on a spin network that has only a single point
of intersection with $\Sigma$. Using the same notation as above, we obtain
\begin{eqnarray}
   \lefteqn{\hat{E}^{\,2}(\Sigma) \, \Psi_S(A)} \hspace{0.5cm} \nonumber \\
   & & = - \Big[ R^j \big( U[A,\gamma_1]\big) \, {}^{\scriptscriptstyle (j)}\tau^i \:
   {}^{\scriptscriptstyle (j)}\tau^i  \,
   R^j \big( U[A,\gamma_2]\big) \Big]_{mn} \, \Psi^{mn}_{S-\gamma}(A) \nonumber \\
   & & = \Big[ R^j \big( U[A,\gamma_1]\big) \, j(j+1) \,
   R^j \big( U[A,\gamma_2]\big) \Big]_{mn} \, \Psi^{mn}_{S-\gamma}(A) \nonumber \\ 
   & & =  j(j+1) \, \Big[ R^j \big( U[A,\gamma_1]\big) \,
   R^j \big( U[A,\gamma_2]\big) \Big]_{mn} \, \Psi^{mn}_{S-\gamma}(A) \nonumber \\ 
   & & = j(j+1) \, \Psi_S(A) \;.
\label{E2onPsi}
\end{eqnarray}
Here $\mathcal{C} := -{}^{\scriptscriptstyle (j)}\tau^i \: 
{}^{\scriptscriptstyle (j)}\tau^i = j(j+1) \times \bb1$ is the Casimir 
operator of $SU(2)$.  

Thus it seems we are lucky this time. We have found the important result that 
the spin network state is an eigenstate of this seemingly gauge invariant operator 
and even calculated its eigenvalues. 
But so far we have calculated this result only in case of a \emph{single} 
intersection between $S$ and $\Sigma$.
It is easy to convince oneself that for \emph{several} points of intersection
crossterms would appear that again spoil the gauge invariance of 
$\hat{E}^{\,2}(\Sigma) \,\Psi_S(A)$. However, using a simple trick, these 
crossterms may be eliminated in order to construct a genuinely 
$SU(2)$ gauge invariant operator in the following way.

\begin{figure}
\centerline{\includegraphics[width=7.5cm]{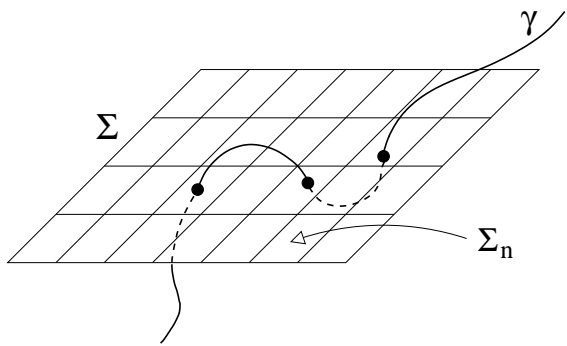}}
\caption[]{A partition of $\Sigma$.}
\label{partition}
\end{figure}

Since we have shown that in the case of a \emph{single} intersection 
$\hat{E}^{\,2}(\Sigma)$ turns out to be an operator of the type 
we are looking for, it is natural to consider a partition $\rho$ of $\Sigma$ into
$n$ small surfaces $\Sigma_n$, where $\bigcup_n \Sigma_n = \Sigma$, 
in such a way that for any given spin 
network $S$ all different points of intersection $P$ lie in distinct surfaces 
$\Sigma_n$. This is shown in Fig.~\ref{partition} for a curve $\gamma$ which
intersects the surface several times.
Clearly $n = n(\rho)$ depends on the degree of refinement of the partition.

\begin{figure}  
\centerline{\includegraphics[width=7cm]{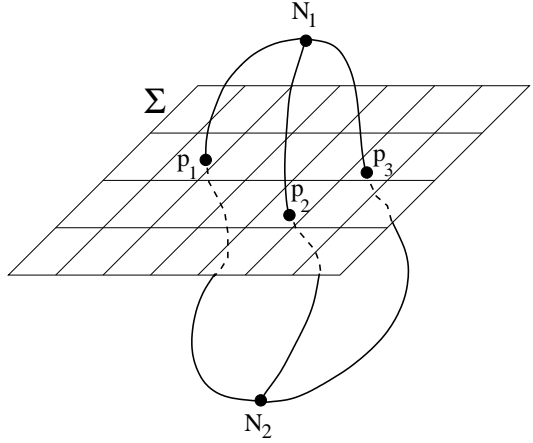}}
\caption[]{A simple spin network $S$ intersecting the surface $\Sigma$.}
\label{spinnet_on_sigma}
\end{figure}

Hence we obtain a new operator $\widehat{\mathfrak A}(\Sigma)$ which is defined in the 
limit of infinitely fine
triangulations or partitions of $\Sigma$, respectively, as
\begin{equation}
\label{area_op}
  \widehat{\mathfrak A}(\Sigma) := \lim_{\rho \to \infty} \, \sum_{n = n(\rho)} 
        \sqrt{\hat{E}^i (\Sigma_n) \, \hat{E}^i (\Sigma_n)} \;.
\end{equation}
The square root is introduced for later convenience.
It can be shown easily that this operator is defined independently of the
partition $\rho$ chosen.  
For simplicity, we disregard spin networks that have either a node lying on $\Sigma$ 
or a continuous, i.e. infinite (non-countable) number of intersection points with 
it, cf. Fig.~\ref{spinnet_on_sigma}.

\begin{figure}[b]
\centerline{\includegraphics[width=6cm]{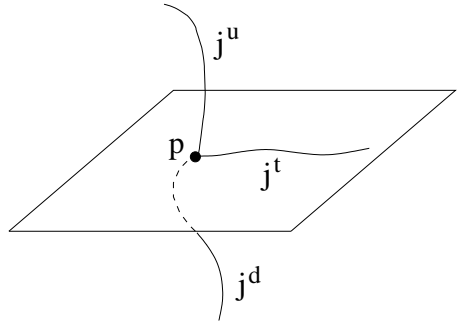}}
\caption[]{The three classes of links that meet in a node on the surface.}
\label{node_on_sigma}
\end{figure}

Then, using (\ref{E2onPsi}) we obtain immediately the action of 
$\widehat{\mathfrak A}(\Sigma)$ on a spin network state as
\begin{equation}
  \widehat{\mathfrak A}(\Sigma) \, \Psi_S(A) = \sum_{P \, \in \, S \, \cap \, \Sigma} 
        \sqrt{j_P(j_P+1)} \: \Psi_S(A) \;.
\end{equation}
Hence, each link of the spin network $S$ labelled by the irreducible representation
$j$ of $SU(2)$ which crosses the surface transversely in the small surface 
$\Sigma_n$ contributes a factor of $\sqrt{j(j+1)}$. 
Other subsurfaces $\Sigma_{n'}$ that have no intersection with a link 
of $S$ give no contribution.
Since the operator is diagonal on spin network states and real on this basis,
it is also self-adjoint.

To summarize, we have obtained for each surface $\Sigma \in M$ a 
well-defined $SU(2)$ gauge invariant and self-adjoint 
operator $\widehat{\mathfrak A}(\Sigma)$, 
which is diagonalized in the spin network basis on 
$\mathcal{H}_0$, the Hilbert space of gauge invariant state functionals.
The corresponding spectrum (with the restrictions mentioned) is labeled by
multiplets $\vec{j} = (j_1, \ldots , j_n)$, $i=1, \ldots, n$, and $n$ arbitrary, of 
positive half integers $j_i$. This is called \emph{main sequence} of the spectrum and
is given (up to constant factors) by
\begin{equation}
\label{mainsequence}
  {\mathfrak A}_{\vec{j}}(\Sigma) = \sum_{i} \sqrt{j_i(j_i+1)} \;.
\end{equation}

As already mentioned, (\ref{mainsequence}) is not the result of the
most general case, since we excluded crossings of $S$ and 
$\Sigma$, in which the intersection points $P$ may be nodes $p$ of the spin network.
To complete the picture and include these cases, 
we finally give the full spectrum of $\widehat{\mathfrak A}(\Sigma)$, 
which was calculated in \cite{FritelliLehnerRovelli96} directly
in the loop representation and in \cite{AshtekarLewand97} in the connection representation.
In the general case we may divide the links that meet at the nodes on the surface 
into three classes according to their relative 
position with respect to the surface, see Fig.~\ref{node_on_sigma}. First, there are
the \lq \lq tangential'' $(t)$ links which lie entirely in $\Sigma$. The remaining
two classes are given by the \lq \lq up'' $(u)$ and \lq \lq down'' $(d)$ 
links according to the (arbitrary) side of $\Sigma$ they lie on. 

The \emph{full} spectrum of (\ref{area_op}) is labeled by $n$-tuplets of triplets of positive 
half integers $j_i$, namely
$\vec{j}_i = (j^u_i, j^d_i, j^t_i)$, $i=1, \ldots, n$, and $n$ arbitrary.
It is given by
\begin{equation}
\label{scndsequence}
  {\mathfrak A}_{\vec{j}_i}(\Sigma) = \frac{1}{2} \,
  \sum_{i} \sqrt{2 j^u_i (j^u_i +1) + 2 j^d_i (j^d_i +1) - j^t_i (j^t_i +1)}\;.
\end{equation}
It contains of course the previous case (\ref{mainsequence}) which corresponds to
the choice $j^u_i = j^d_i$ and $j^t_i=0$.
Those eigenvalues which are contained in (\ref{scndsequence}) but not in 
(\ref{mainsequence}) are called the \emph{second sequence}.

%%%%%%%%%%%%%%%%%%%%%%%%%%%%%%%%%%%%%%%%%%%%%%%%%%%%%%%%%%%%%%%%%%%%%%%%%%%%%%%%%%%%%%%
%%%%%%%%%                              Section 3                              %%%%%%%%% 
%%%%%%%%%%%%%%%%%%%%%%%%%%%%%%%%%%%%%%%%%%%%%%%%%%%%%%%%%%%%%%%%%%%%%%%%%%%%%%%%%%%%%%%

\section{Quantization of the Area}
\label{sect_area}

In the previous section we have described the 
construction and diagonalization of the $SU(2)$ gauge invariant and 
self-adjoint operator $\widehat{\mathfrak A}(\Sigma)$ using a basis of spin network states in 
the kinematical gauge invariant Hilbert space $\mathcal{H}_0$.
So far, the physical interpretation of this operator
was totally disregarded.
Explicitely, we have studied the operator
\begin{equation}
\label{area_op2}
  \widehat{\mathfrak A}(\Sigma) := \lim_{\rho \to \infty} \, \sum_{n = n(\rho)} 
        \sqrt{\hat{E}^i (\Sigma_n) \, \hat{E}^i (\Sigma_n)} \;.
\end{equation}
In the following, we are going to look for the corresponding classical quantity.
Just as in usual quantum mechanics this amounts to replacing the 
quantum operators $\hat{E}^i(\Sigma)$ by their classical analogues.

The conjugate momentum operator, which essentially is given by 
\mbox{$\delta / \delta A^i_a(x)$}, is the quantum analogue of the (smooth) inverse 
densitized triad $E^a_i(x)$, i.e. we have the correspondence 
\begin{equation}
\label{correspond} 
   E^a_i(x) \; \longleftrightarrow \; -\, i \hbar G \, \frac{\delta}{\delta A^i_a (x)} 
\end{equation}
between classical and quantum quantities,
as we already stated in sect.~\ref{ops_on_H}. 
We replace the operator (\ref{area_op2}) in the classical limit
by its analogue (\ref{correspond}),
\begin{equation}
\label{classical_area}
  {\mathfrak A}(\Sigma) := \lim_{\rho \to \infty} \, \sum_{n = n(\rho)} 
  \sqrt{E^i (\Sigma_n) E^i (\Sigma_n)} \;,
\end{equation}
and study its physical meaning.
As before, we use again $\hbar = G = 1$. Moreover,
\begin{equation}
\label{E_classical}
  E^i (\Sigma_n) = \int_{\Sigma_n} d\sigma^1 d\sigma^2 \, n_a (\vec{\sigma}) \,
        E^{ia}\big( \vec{x}(\vec{\sigma}) \big)
\end{equation}
is the classical analogue of the smeared version (\ref{E_operator}) of the  
operator $\hat{E}^i (\Sigma_n)$ defined on one specific subsurface $\Sigma_n$ of the triangulation 
$\rho$ of $\Sigma$, and
\begin{equation}
  n_a(\vec{\sigma}) = \epsilon_{abc} \, \frac{\partial x^b(\vec{\sigma})}{\partial \sigma^1}
        \frac{\partial x^c(\vec{\sigma})}{\partial \sigma^2}        
\end{equation}
is the normal to $\Sigma_n$.
For a sufficiently fine partition $\rho$, i.e. arbitrarily small surfaces $\Sigma_n$, the 
integral (\ref{E_classical}) can be approximated by
\begin{equation}
  E^i (\Sigma_n) \approx \Delta \sigma^1 \Delta \sigma^2 \, n_a(\vec{\sigma})\, 
  E^{ai}\big(\vec{x}_n (\vec{\sigma})\big),
\end{equation}
where $\vec{x}_n$ is an arbitrary point in $\Sigma_n$ and 
$\left(\Delta \sigma^1 \Delta \sigma^2 \right)$
denotes its coordinate area. 
Inserting this result back into the classical expression (\ref{classical_area})
gives
\begin{eqnarray}
  \label{riemann_int1}
  {\mathfrak A}(\Sigma) &=& \lim_{\rho \to \infty} \, \sum_{n = n(\rho)}\! \Delta \sigma^1 
	\Delta \sigma^2 
  \sqrt{n_a(\vec{\sigma}) E^{ai}\big(\vec{x}_n (\vec{\sigma})\big) \, 
  n_b(\vec{\sigma}) E^{bi}\big(\vec{x}_n (\vec{\sigma})\big)} \\
  \label{riemann_int2}
  &=& \int_{\Sigma} d^2\sigma
  \sqrt{n_a(\vec{\sigma}) E^{ai}\big(\vec{x} (\vec{\sigma})\big) \, 
  n_b(\vec{\sigma}) E^{bi}\big(\vec{x} (\vec{\sigma})\big)} \;.
\end{eqnarray}
The second line (\ref{riemann_int2}) follows immediately by noting that 
(\ref{riemann_int1}) is nothing but the definition of the Riemann integral.
For its evaluation we choose local coordinates in such a way that $x^3(\vec{\sigma})=0$ 
on $\Sigma$ and furthermore $x^1(\vec{\sigma})=\sigma^1$, $x^2(\vec{\sigma})=\sigma^2$, 
resulting in $n_a = n_b = (0,0,1)$.
We obtain 
\begin{eqnarray}
  {\mathfrak A}(\Sigma) &=& \int_{\Sigma} d^2\sigma \sqrt{E^{3i}(\vec{x})\, 
  E^{3i}(\vec{x})} \\
  \label{area2}
  &=& \int_{\Sigma} d^2\sigma \sqrt{ \det g(\vec{x}) \, g^{33}\vec{x}} \\
  \label{area3}
  &=& \int_{\Sigma} d^2\sigma \sqrt{g_{11} g_{22} - g_{12} g_{21}}\\
  \label{area4}
  &=& \int_{\Sigma} d^2\sigma \sqrt{\det \left( {}^2g\right)} \;.
\end{eqnarray}
For the derivation of (\ref{area2}) we used the relation 
between the 3-metric and the triad variables, which is 
$g^{ab}(\vec{x}) \det g(\vec{x}) = E^{ai}(\vec{x}) E^{bi}(\vec{x})$.
The transition to the next equation is made by using the definition for the inverse
of a matrix. 
Noting that $\left({}^2g \right)$ is the 2-dimensional metric induced by 
$g_{ab}$ on $\Sigma$, one recognizes the result (\ref{area4}) as the 
covariant expression for the \emph{area} of $\Sigma$.

In fact, since the classical geometrical observable \lq \lq area of a surface'' 
is a functional of the metric, i.e. of the
gravitational field, in a quantum theory of gravity, where the metric is an 
operator, the area turns into an operator as well.
If this operator reveals a discrete spectrum, this would, according to
quantum mechanics, also imply the discreteness of physical areas at the Planck length.
Recalling the result we obtained in the last section, this means that
the area is quantized!

Restoring physical units and the neglected Immirzi parameter $\beta$,
we get the (main sequence of) eigenvalues of the area as
\begin{equation}
\label{eigenvalues}
  {\mathfrak A} (\Sigma)= 8 \pi \beta \hbar G \sum_{i} \sqrt{j_i(j_i+1)} \;.
\end{equation}
Here we use the notation and results we 
obtained in sect.~\ref{ops_on_H}, namely the discreteness of the spectrum of 
$\widehat{\mathfrak A}(\Sigma)$, which from now on will be denoted as \emph{area operator}.
The formula (\ref{eigenvalues}) gives the area of a surface $\Sigma$ that is 
intersected by a spin network $S$ without having nodes lying in it. 
The quanta are labeled by multiplets $\vec{j}$ of half integers as already realized in
sect.~\ref{ops_on_H}.
The generalization to the case where nodes of $S$ are allowed to lie in $\Sigma$, 
yielding the full sequence of eigenvalues, is given by (\ref{scndsequence}).

%%%%%%%%%%%%%%%%%%%%%%%%%%%%%%%%%%%%%%%%%%%%%%%%%%%%%%%%%%%%%%%%%%%%%%%%%%%%%%%%%%%%%%%
%%%%%%%%%                              Section 4                              %%%%%%%%%
%%%%%%%%%%%%%%%%%%%%%%%%%%%%%%%%%%%%%%%%%%%%%%%%%%%%%%%%%%%%%%%%%%%%%%%%%%%%%%%%%%%%%%%

\vspace*{0.5cm}
\section{The Physical Contents of Quantum Gravity and the \\ Meaning of Diffeomorphism 
Invariance}
\label{sect_diffeoInv}
Some questions arise immediately from the results we discussed in the last section.
 
\vspace{2.5mm}
\emph{Is ${\mathfrak A}(\Sigma)$ observable in quantum gravity?} 

\vspace{2mm}
\noindent and, in general:

\vspace{2mm}
\emph{What should a quantum theory of gravitation predict?}
\vspace{2.5mm}

\noindent These questions are intimately related to the issue of observability 
in both classical and quantum gravity---an issue which is far from being trivial.
Let us begin with an examination of the classical theory.
For a closer look at this topic, we refer to 
\cite{Rovelli91b,Rovelli97c,Rovelli99}.

\subsection{Passive and Active Diffeomorphism Invariance}
We consider ordinary classical general relativity formulated on a
4-di\-men\-sio\-nal manifold $\mathcal{M}$ on which we introduce local coordinates
$x^{\mu},\, \mu = 0, \ldots, 3$, abbreviated by $x$. 

The Einstein equations
\begin{equation} 
\label{eequat}
  R_{\mu\nu} - \frac{1}{2} \, R \, g_{\mu\nu} = 0
\end{equation}
are invariant under the
group of 4-dimensional diffeomorphisms $Diff(\mathcal{M})$ of $\mathcal{M}$. 
Recall that a diffeomorphism $\phi$ is a $C^\infty$ map 
between manifolds that is one-to-one, onto and has a $C^\infty$ inverse. In
other words, the diffeomorphism group is formed by the set of mappings 
$\phi: \mathcal{M} \rightarrow \mathcal{M}$ which preserve the structure of 
$\mathcal{M}$. We consider diffeomorphisms which are given in local coordinates 
by the smooth maps
\begin{equation}
  f \; : \; x^{\mu} \longrightarrow f^{\mu}(x) \;.
\end{equation}
Suppose a solution $g_{\mu\nu}(x)$ of Einstein's equations (\ref{eequat}) is given, 
then due to diffeomorphism invariance, $\tilde{g}_{\mu\nu}(x)$ is also a solution,
where
\begin{equation}
\label{metric_trafo}
\tilde{g}_{\mu\nu}(x) = 
        \frac{\partial f{^\rho}(x)}{\partial x{^\mu}} 
        \frac{\partial f{^\sigma}(x)}{\partial x{^\nu}}
        \: g_{\rho\sigma}\big(f(x)\big) \;.
\end{equation}
There are two geometrical interpretations of (\ref{metric_trafo}) commonly
known as \emph{passive} and \emph{active} diffeomorphisms.

Passive diffeomorphism invariance refers to invariance under change of coordinates,
i.e. the same object is represented in different coordinate systems. 
Choose a (local) co\-or\-di\-nate system $S$ in which the metric is 
$g_{\mu\nu}(x)$. In a second system $S'$
the metric is given by $\tilde{g}_{\mu\nu}\left(f(x)\right)$. Satisfying (\ref{metric_trafo}),  
both of them represent the \emph{same} metric on $\mathcal{M}$.

Active diffeomorphisms on the other hand relate different objects in $\mathcal{M}$ 
in the same coordinate system. This means that 
$f$ is viewed as a map associating one point in the manifold to another one.
Take for example two points $P, Q \in \mathcal{M}$ and consider 
two metrics $g_{\mu\nu}(x)$ and $\tilde{g}_{\mu\nu}(x)$, which are both
solutions of (\ref{eequat}).
Then the distance $d$ between $P$ and $Q$ computed using the two metrics is
different, i.e.
$d_{g}(P,Q) \neq d_{\tilde{g}}(P,Q)$. We have two \emph{distinct}
metrics on $\mathcal{M}$ which both solve Einstein's equations.
These two metrics might still be related by equation (\ref{metric_trafo}), i.e. they
are related by an active diffeomorphism.

\begin{figure}
\centerline{\includegraphics[width=11cm]{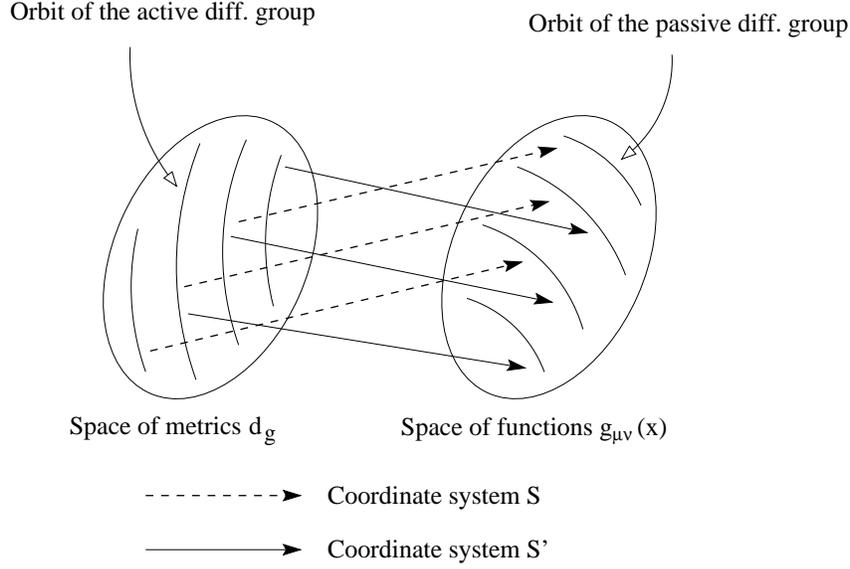}}
\caption[]{The relation between active and passive diffs and the choice of 
coordinates.} 
\label{diffeos}
\end{figure}

The relations between active and passive diffeomorphisms, as well as the choice 
of coordinates, is clarified in Fig.~\ref{diffeos}.

In order to avoid confusion with regard to passive and active diffeomorphisms
in coordinate-dependent considerations, we simply drop coordinates and pass over to the 
coordinate-free formulation.
Thus we consider the manifold $\mathcal{M}$ with metric $g$,
defined as the map
\begin{eqnarray}
  g: \mathcal{M} \times \mathcal{M} &\rightarrow& \mathbb{R} \\
  \left( P , Q \right) &\mapsto& d_g \left(P,Q\right) \;,
\end{eqnarray}
where $P, Q \in \mathcal{M}$. Suppose $d_g$ solves Einstein's 
equations.  A diffeomorphism $\phi: \mathcal{M} \rightarrow 
\mathcal{M}$ acts as a smooth displacement over the manifold, 
resulting in $d_{\tilde{g}}$, 
\begin{equation}
d_{\tilde{g}} \left(P,Q\right) = d_{g} 
\left(\phi^{-1}(P),\phi^{-1}(Q)\right) \; . 
\end{equation}
Active diffeomorphism invariance is the fact that if $d_{g}$ is 
a solution of the Einstein theory, so is $d_{\tilde{g}}$.  This 
shows that Einstein's theory is invariant under (active!)  
diffeomorphisms even in a coordinate free formulation.

General relativity is distinguished from other dynamical field 
theories by its invariance under \emph{active} diffeomorphisms.  
Any theory can be made invariant under \emph{passive} 
diffeomorphisms.  Passive diffeomorphism invariance is a 
property of the {\em formulation\/} of a dynamical theory, 
while active diffeomorphism invariance is a property of the 
dynamical theory \mbox{{\em itself}}. Invariance under smooth 
displacements of the \emph{dynamical} fields holds only in 
general relativity and in any general relativistic theory.  It 
does not hold in QED, QCD, or any other theory on a fixed (flat 
or curved) background.

\subsection{Dirac Observables}
\begin{figure}[b]
\centerline{\includegraphics[width=4cm]{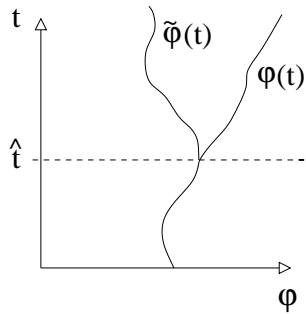}}
\caption[]{An example for a not uniquely determined evolution of a dynamical system.}
\label{evolution}
\end{figure}

Consider a classical dynamical system whose equations of motion do not uniquely
determine its evolution, as pictorially illustrated in Fig.~\ref{evolution}.
The two solutions $\varphi\,(t)$ and 
$\tilde{\varphi}\,(t)$ which evolve from the same set of initial data,
separate at some later time $\hat{t}$, i.e. 
\begin{eqnarray}
  \varphi\,(t) &=& \tilde{\varphi}\,(t) \hspace{6mm} \mbox{if}  \hspace{6mm} t < \hat{t}\\
  \varphi\,(t) &\neq& \tilde{\varphi}\,(t) \hspace{6mm} \mbox{if}  
  \hspace{6mm} t \geq \hat{t} \geq 0 \;.
\end{eqnarray}

\noindent Then, as first accurately argued by Dirac, $\varphi\,(t)$ and $\tilde{\varphi}\,(t)$ 
must be physically 
indistinguishable or \emph{gauge-related}, respectively. Otherwise determinism, which is 
a basic principle in classical physics, would be lost.
Dirac gave the definition of observables respecting determinism in the following way.
A \emph{gauge invariant} or \emph{Dirac observable} is a function $\mathcal{O}$ of 
the dynamical variables that does not distinguish 
$\varphi\,(t)$ and $\tilde{\varphi}\,(t)$, i.e.
\begin{equation}
  \mathcal{O}\big(\varphi\,(t)\big) = \mathcal{O}\big(\tilde{\varphi}\,(t)\big) \;.
\end{equation}
In other words, only those observables that have the same values on the solutions
$\varphi\,(t)$ and $\tilde{\varphi}\,(t)$ can be observed.
Hence the theory can predict only Dirac observables.

Does this imply that any physical quantity that we measure is necessarily 
a Dirac observable? It turns out that the answer is in the negative.
To understand this sublety, consider the example of a simple pendulum described by the 
variable $\alpha$ which 
is the deflection angle out of equilibrium. The motion of the pendulum is given by
the evolution of $\alpha$ in time $t$, namely by $\alpha(t)$. Since
$\alpha(t)$ is predicted by the equation of motion for any time $t$ once the 
initial data set is fixed, it is a Dirac observable. One should notice that we
are actually describing a system in terms of \emph{two} physical quantities 
rather than one, namely  
the pendulum itself, described by position $\alpha$, and a clock measuring the 
time $t$. However, in contrast to position at a given time, there is no way how 
time itself could be \lq \lq predicted''. It simply tells us \lq \lq when'' we are. 
Therefore, $t$ is a measureable quantity but it is not a Dirac observable. To state 
this more precisely, we introduce the notion of \emph{partial observables}.
We call $t$ an \emph{independent partial observable} and $\alpha$ a 
\emph{dependent partial observable}. The Dirac observable is given by $\alpha(t)$.

There is an important relation between Dirac observables 
and the Hamitonian formalism. Dirac observables are characterized by having vanishing 
Poisson brackets with the constraints.
In fact, the entire constrained system formalism was built by 
Dirac with the purpose of characterizing the gauge invariant or
Dirac observables, respectively. 
To elucidate this feature, consider a classical dynamical system with 
canonical Hamiltonian $H_0$, as well as $k$ additional constraints 
\begin{equation}
  C_m = 0 \; ,\hspace{6mm} m=1, \dots, k \;, 
\end{equation}
defined on phase space. 
The complete Hamiltonian, which is defined on the full phase space, 
is then given by 
\begin{equation}
  H = H_0 + N_m(t)\, C_m \;,
\end{equation}
with $k$ arbitrary functions $N_m(t)$. The dynamics of an observable $\mathcal{O}$ 
is given by the Hamilton's equations
\begin{equation}
  \frac{d\mathcal{O}}{dt} = \{ \mathcal{O}, H\} + N^m(t) \{\mathcal{O}, C_m \} \;.
\end{equation}
From this one recognizes immediately that the evolution is deterministic, and thus 
$\mathcal{O}$ a Dirac observable, only if 
\begin{equation}
  \{\mathcal{O}, C_m \} = 0 \hspace{1cm} \forall \; m \; ,
\end{equation}
just as claimed before.

\subsection{The Hole Argument}
Dirac's postulate that only gauge invariant or Dirac observables, respectively, 
can be measurable quantities, was applied to general relativity by Einstein himself
in his famous \lq \lq hole argument'' from 1912, cf. \cite{EinsteinGrossmann14}.

\begin{figure}
\centerline{\includegraphics[width=7cm]{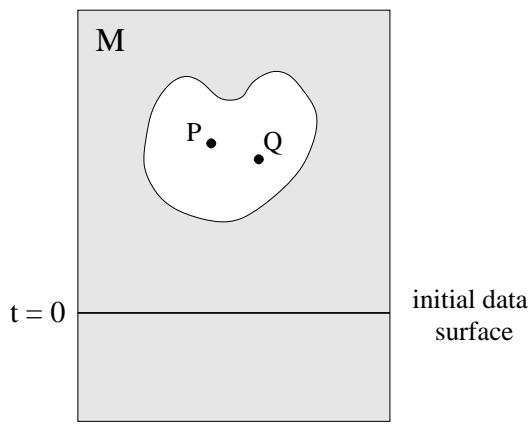}}
\caption[]{The hole argument.}
\label{hole}
\end{figure}

Suppose we have a space--time $\mathcal{M}$ including other structures that represent
matter (e.g. scalar fields or particles). Suppose furthermore that the matter configuration 
is such that
there is a \emph{hole} in space--time, i.e. a region without matter, as indicated in 
Fig.~\ref{hole}. Let
$g_{\mu\nu}(x)$ and $\tilde{g}_{\mu\nu}(x)$ be two distinct metrics which are equal 
everywhere in $\mathcal{M}$ except for the hole, but nevertheless, both are supposed to
solve Einstein's equations.
Now we introduce a spacelike (initial data) surface such that the hole is entirely in the
future of it. Since the metrics are equal everywhere outside, they do have the 
same set of initial data on the surface. 

If we now consider the distance $d$ between two distinct points $P$ and $Q$ which  
are both inside the hole, we note immediately that $d_g(P,Q) \neq d_{\tilde{g}}(P,Q)$, 
although the metrics have the same inital conditions.
Hence, according to the discussion in the previous section, $d_g$ is 
\emph{not} a Dirac observable.
So it seems that we uncovered a mystery of the theory! 
The distance is not an observable predicted by the theory.
Then the obvious question we have to ask is: 

\vspace{2.5mm}
\lq \lq \emph{What} is predicted by general relativity at all?''
\vspace{2.5mm}

\begin{figure}[t]
\centerline{\includegraphics[width=7cm]{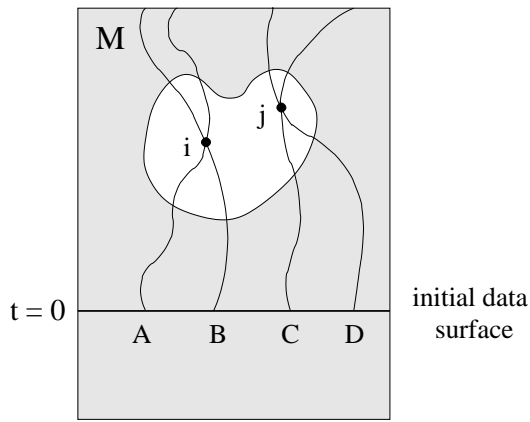}}
\caption[]{A solution to the hole argument.}
\label{fourparticles}
\end{figure}

\noindent Einstein was so impressed by this conclusion, that he claimed 
in 1912 that general covariance could {\em not\/} be a property 
of the theory of gravity.  
It took some time---three years---until Einstein presented the 
solution to this puzzle, and thus got back to general covariance,
in 1915. To illustrate his strategy, we consider a setting similar to
the one above, which corresponds to Fig.~\ref{hole}.
More precisely, we consider general relativity and 4 particles denoted 
as $A, B, C$ and $D$.
Their trajectories are determined by the equations of motion and they
are supposed to start at the spacelike inital surface, as shown in
Fig.~\ref{fourparticles}. 
Furthermore, we suppose that $A$ and $B$ meet in $i$ inside the hole,
and $C$ and $D$ meet in $j$ inside the hole as well.
Consider now the distance $d$ between the point $i$ and the 
point $j$.  Is $d$ a Dirac observable?  At first sight, we are 
in the same situation as above, but there is an essential 
subtle difference in the way we have defined the observable.  
Consider now the diffeomorphism that 
sends $g_{\mu\nu}(x)$ into $\tilde{g}_{\mu\nu}(x)$. 
Since the theory is invariant only under a diffeomorphism that acts on 
{\em all\/} its dynamical variables, $\tilde{g}_{\mu\nu}(x)$ 
is a solution of the Einstein equations only if the 
diffeomorphism displaces the trajectories of the particles as 
well.  Thus $i$ and $j$ will also be displaced by the diffeomorphism.
Then, after having performed the active diffeomorphism, the new distance 
between the intersection points is
\begin{equation}
  \tilde d=d_{\tilde g}\big(\phi(i),\phi(j)\big) = 
  d_{g}\big(\phi^{-1}\phi(i),\phi^{-1}\phi(j)\big) = d_{g}(i,j) = d\; . 
\end{equation}
Hence it follows that this distance is gauge invariant. The distance $d$ 
between the intersection points is indeed a Dirac observable.

One can extend this setting also to cases which involve fields. 
As an example, consider general relativity and 2 additional fields, 
namely $g_{\mu\nu}(x)$, $\varphi_t(x)$, and $\varphi_z(x)$.
Then the area $\mathfrak A$ of the $\varphi_t = \varphi_z = 0$ surface is a Dirac 
observable as well, and is given by
\begin{equation}
  {\mathfrak A} = \int_{\varphi_t=0 \atop \varphi_z=0} d^2\sigma \, \sqrt{\det {}^2g} \;.
\end{equation}

As a slightly generalized example, consider general relativity and  
three fields, i.e. $g_{\mu\nu}(x)$, $\varphi_t(x)$,  $\varphi_z(x)$, and  
$\varphi_{\scriptscriptstyle\Sigma}(x)$, the area ${\mathfrak A}(\Sigma)$ of the 
surface determined by
\begin{equation}
  \varphi_t = \varphi_z = 0,\;\; \mbox{and}\;\; \varphi_{\scriptscriptstyle \Sigma} \geq 0 
\end{equation}
is given by
\begin{equation}
  {\mathfrak A}(\Sigma) = \int_{\varphi_t=0 \atop \varphi_z=0} d^2\sigma \, 
        \delta(\varphi_{\scriptscriptstyle\Sigma})
        \, \sqrt{\det {}^2g} \;.
\end{equation}
The reader can convince herself that a diffeomorphism transforming all the fields
\linebreak
$\big(g_{\mu\nu}(x), \varphi_t(x), \varphi_z(x), \varphi_{\scriptscriptstyle\Sigma}(x)\big)$
does not change the number ${\mathfrak A}(\Sigma)$. Thus ${\mathfrak A}(\Sigma)$ is a
Dirac observable.

In general, to define \lq \lq local'' Dirac observables in general 
relativity we have to use some of the degrees of freedom of the 
theory (the particles, the fields) for localizing a space--time 
point or a space--time region.  It is important to notice that in 
principle we do not need \emph{matter} or fields to do so.  
Instead, we can use part of the degrees of freedom of the 
graviational field itself.  This strategy was followed for 
instance by Komar and Bergman by defining 4 curvature scalars 
and using them as physically defined coordinates \cite{BergmannKomar72}. 
While formally correct, the use of gravitational degrees of freedom 
for defining observables in general relativity leads us far 
away from observables concretely used in realistic 
applications of general relativity, all of which use matter 
degrees of freedom for localizing the observables.  An example 
of a realistic observable used in physical applications of 
general relativity is the physical distance between two 
space--time events, one on a Global Positioning System (GPS) satellite 
and one on a Earth based GPS station.  In this case, matter 
degrees of freedom (coupled to gravity) localize two space--time 
points and the distance between them is a Dirac observable.

To sum up, we have seen that the puzzle of the hole argument
can be resolved. Physical quantities predicted by general relativity, i.e.
Dirac observables, can be defined inside the hole. But in order to 
\lq \lq localize'' points, we have to use some dynamical quantity.
The most realistic way of doing so is to use matter.
In other words, Dirac observables are defined in space--time regions
which are determined by dynamical objects.

In the following section, we will see that this definition of 
localization, which is necessary in general relativity, implies 
a profound change of our notions of space and time.

\subsection{The Physical Interpretation}
Before considering the conceptual changes in the notions of space and time
brought by general relativity,
it is instructive to reflect on the main modifications that these concepts have 
undergone in the historical development of physics.
The key developments in this business are related to the names of Descartes (and Aristotle), 
Newton, and last but not least, Einstein.

According to Descartes, there is no \lq \lq space'' at all, but only physical 
objects which can be in touch with each other.
The \lq \lq position'' or location, respectively, of an object is only 
defined by the naming of other physical objects close to it, i.e. the position of a body 
is the set of those objects to which the body is contiguous. 
Equally important is the concept of \lq \lq motion'', which is defined as the 
change of position.
Thus motion is determined by the change of contiguity, i.e. only in relation to
other objects. This point of view is denoted as \emph{relationalism}.
Descartes' definitions of space, position, and motion are by the way essentially 
the same that were given by Aristotle.

An important historical step was then provided by Newton's definition of physical space.
According to Newton, \lq \lq space'' exists by itself, independently of the objects in it.
Motion of a body can be defined with respect to space alone, 
irrespectively whether other objects are present.  Newton 
insists on this points, on the ground that acceleration can be 
defined absolutely.  In fact, it is only thanks to the fact 
that acceleration is defined in absolute terms, that the entire 
structure of Newton's mechanics ($F=ma$) holds.  Newton 
discussed the fact that acceleration is absolute in the famous 
example of the rotating bucket, which shows that the absolute
rotation of the water, and not the rotation with respect to the 
bucket, has observable consequences.  Thus, according to 
Newton, space exists independently of objects, weather they are 
present or not.  The location of objects is the part of space 
that they occupy.  This implies that motion can be understood 
without regard to surrounding objects.  Similarly, Newton uses 
absolute time, leading to a space--time picture which provides 
an always present fixed background over which physics takes 
place.  Objects can always be localized in space and time with 
respect to this fixed non-dynamical background.

But if there is \lq \lq space'' which is always present, how can it be 
captured, or observed? This can be done by using reference systems.
The great idea was to select some physical bodies (like walls, rules or clocks)
and treat them as reference systems.
Physically one has to distinguish the dynamical objects that one wants to study 
from reference system objects. They are dynamically decoupled. 

In the language introduced earlier, the dynamical objects define dependent partial 
observables, while the objects referred to as reference system define
independent partial observables.
Examples of dynamical objects may be the deflection angle $\alpha$ of a pendulum, or 
the position $\vec{x}$ of a particle. An example of a reference system variable is 
the clock time $t$.
The Dirac observables would then just be $\alpha(t)$ and $\vec{x}(t)$.

As an example, we consider the case of a pendulum.
The differential equation governing this dynamical system is (for small oscillations) 
just 
\mbox{$\ddot{\alpha}(t) = - \omega^2 \, \alpha(t)$}. The solution is
\begin{equation}
\label{pend}
  \alpha(t) = A \sin (\omega t + \varphi) \;.
\end{equation}
A {\em state\/} is determined by the constants $A$ and 
$\varphi$, or equivalently, by initial position and velocity at 
some fixed time.  Once the state (i.e. $A$ and $\varphi$) is known, 
the functional dependence $\alpha(t)$ between the dependent and 
independent observables can be computed.  In fact, it is given 
by (\ref{pend}).

Thus, in the Newtonian scheme, we have a fixed space and a 
fixed time, revealed by the objects of the reference system.  
The objects forming the reference system determine localization 
in space and in time and define partial observables ($t$, above) 
which are not dynamical variables in the dynamical models one 
considers.

In general relativity things change profoundly.
We have seen in the discussion of the hole argument and its solution,
that the theory does not distinguish reference system objects from dynamical 
objects. This means that independent and dependent physical 
observables are not distinguished any more!
The reference system can not be decoupled from the dynamics.
Therefore, in the Einsteinian framework the notion of \lq \lq dynamical object'' 
has to be extended compared to the Newtonian case, since now also the reference 
system objects are included as dynamical variables.
Localization of observables is determined by other variables of the theory.
Therefore:

\vspace{2mm}
\emph{Position and Motion are fully relational in General Relativity!} 
\vspace{2mm}

\noindent This important statement is the same as provided in the Cartesian--Aristote\-lian 
picture.

The essential consequence of the fact that localization of dynamical objects 
in general relativity is defined only with respect to each other, 
is the appearance of the diffeomorphism group.
Indeed, if we displace all \emph{dynamical} objects in 
the manifold at once, we generate nothing but an equivalent 
mathematical description of the same physical state, because 
localization with respect to the manifold is irrelevant.  In other words,
the individual mathematical points in the manifold have no intrinsic
physical significance. Only relative localization is relevant.  This is precisely the claim 
of active diffeomorphism invariance of the theory.  Hence, a 
physical state is not located somewhere.  

In a quantum theory of gravity, we should not expect quantum exitations 
\emph{on} space--time, as the Newtonian point of view would imply, rather we should
expect quantum excitations \emph{of} space--time.

The challenge in the construction of quantum gravity is to 
find a quantum field theory in which position and motion are fully relational, i.e.
a quantum field theory without an \emph{a priori} space--time localization.
Here the wheel turns full circle, and we return to loop quantum gravity, which implements
precisely these requirements.

%%%%%%%%%%%%%%%%%%%%%%%%%%%%%%%%%%%%%%%%%%%%%%%%%%%%%%%%%%%%%%%%%%%%%%%%%%%%%%%%%%%%%%%
%%%%%%%%%                              Section 5                              &&&&&&&&&
%%%%%%%%%%%%%%%%%%%%%%%%%%%%%%%%%%%%%%%%%%%%%%%%%%%%%%%%%%%%%%%%%%%%%%%%%%%%%%%%%%%%%%%

\section{Dynamics, True Observables and Spin Foams}
\label{sect_dynamics}
The analysis of the important question of observability in 
general relativity led to the insight that spatiotemporal relationalism \`{a} la
Descartes plays a major role in the formulation of the theory.

In this section we will return to the quantum theory and firstly focus on the 
implementation of relationalism into the framework of canonical quantum gravity. 
Secondly, we will investigate the dynamics and the true, i.e. physical observables
of the theory, which formally amounts to
the still open problem of solving the Hamiltonian constraint. Instead of attacking 
this directely, we will construct 
a projection operator onto the physical states of loop quantum gravity, which will lead 
to a covariant space--time formulation and a relation to the so-called spin foam models.
For a more detailed analysis of this topic we refer to \cite{Rovelli98}.

As we mentioned at the end of the last section, loop quantum gravity is well-suited 
to tackle the matters discussed there.
Thus, the starting point of our considerations is the implementation of the concept of 
\emph{non-localizability} into the framework of loop quantum gravity. 
And of course, as one might have expected, this is achieved by solving the 
diffeomorphism constraint!

Recall from sect.~\ref{su2} that the basis in the gauge invariant Hilbert 
space $\mathcal{H}_0$ is given by the spin network states $\Psi_S(A)$. 
In the following we adopt Dirac's bra-ket notation and denote an abstract basis  
state $\Psi_S$ as $|S\rangle$, such that a state in the connection representation would be 
given by
\begin{equation}
  \Psi_S(A) = \langle A| S \rangle \;.
\end{equation}

\subsection{The Diffeomorphism Constraint}
\label{diffsagain}
In sect.~\ref{subsect_cylfct} we have shown that the Hilbert space $\mathcal{H}$
carries a natural unitary
representation $U(Diff)$ of the diffeomorphism group $Diff(M)$ of the 3-manifold $M$,
\begin{equation}
  U(\phi)\, \Psi(A) = \Psi(\phi^{-1} A)\;, \hspace{1cm} \phi \in Diff(M) \;.
\end{equation}
In the following we will outline the construction of the diffeomorphism 
invariant Hilbert space $\mathcal{H}_{diff}$ (recall Fig.~\ref{construction_of_H}),
which can be considered as the space $\mathcal{H} / Diff(M)$ of solutions of
the quantum diffeomorphism contraint. 

Let us now consider a finite action of a $U(Diff)$ on a spin network state $| S \rangle$.
We get
\begin{equation}
\label{actons}
  U(\phi) | S \rangle = | \phi \cdot S \rangle \;.
\end{equation}
Thus, $U$ sends a state of the spin network basis to another one which is based on a 
shifted graph. To obtain states which are invariant under $U$, one has to solve 
\begin{equation}
\label{diffinvstates}
  U \Psi = \Psi \;.
\end{equation}
However, there is no finite norm state invariant under the action of the 
diffeomorphism group.  This is not surprising, since the gauge 
group is not compact, and leads us to a familiar situation in 
quantum theory.  The way out is to use generalized state 
techniques.  The simplest manner of doing so is (roughly) to solve 
(\ref{diffinvstates}) in $\mathcal{H}^*$, the topological dual of the 
space of finite linear combinations of spin network states.  We 
construct $\mathcal{H}_{diff}$ as the $Diff(M)$ invariant part 
of $\mathcal{H}^*$.

Let $s$ be an equivalence class of embedded spin networks $S$ under the action of
$Diff(M)$, i.e. $S, S' \in s$, if there exists a $\phi \in Diff(M)$, such that 
$S' = \phi \cdot S$. An equivalence class $s$ or abstract spin network, respectively, 
is a spin network
which is \lq \lq smeared'' over $M$. It is usually called $s$-knot.
For each of these $s$-knots an element $\langle s |$ of $\mathcal{H}_{diff}$ is defined.
Since they lie in a subset of the dual of $\mathcal{H}$, they act naturally on spin 
network states as
\begin{equation}
\label{s_scalar1}
  \langle s | S \rangle = \left\{
        \begin{array}{c@{\qquad}l}
                0 & \mbox{if\quad} S \not\in s \\
                \vspace{0mm} & \\
                 1 & \mbox{if\quad} S \in s \;.
        \end{array} \right.   
\end{equation}
A scalar product in $\mathcal{H}_{diff}$ is defined by
\begin{equation}
\label{s_scalar2}
  \langle s | s' \rangle =  c(s) \, \delta_{ss'} \;.
\end{equation}
Here $c(s)$ is the (discrete) number of isomorphisms of an $s$-knot into itself, that 
preserve the coloring and can be obtained from a diffeomorphism of $M$.
One can prove \cite{Marolf} that self-adjoint and diffeomorphism invariant operators in
$\mathcal{H}_0$ are self-adjoint under this inner product when restricted to 
$\mathcal{H}_{diff}$. Thus (\ref{s_scalar2}) is the appropriate physical scalar product 
as we claimed in sect.~\ref{sect_connectionformalism}, picked out by the requirement 
that real classical quantities become self-adjoint operators.
Accordingly, the states $\big(1/\sqrt{c(s)} \big) |s \rangle$ form an orthonormal basis
(notice that we freely interchange bra's and ket's).

The states $| s \rangle$ are the (3d) diffeomorphism invariant quantum states 
of the gravitational field.  They are labelled by abstract, 
non-embedded (knotted, colored) graphs $s$, the \mbox{$s$-knots}. As we have seen 
above, each link of the graph can be seen as carrying a quantum 
of area.  As shown for instance in \cite{DePietriRovelli96}, a 
similar results holds for the volume: in this case, that are 
the nodes that carry quanta of volume.  Thus, an $s$-knot
can be seen as an elementary quantum excitation of space 
formed by \lq \lq chunks'' of space (the nodes) with quantized 
volume, separated by sheets of surface (corresponding to the 
links), with quantized area.  The key point is that an $s$-knot 
does not live on a manifold. The quantized space does reside 
\lq \lq somewhere''. Instead, it defines the \lq \lq where'' by itself.
This is the picture of quantm space--time that emerges from loop 
quantum gravity. 

\subsubsection{Formal Manipulations.}
We close the discussion on the diffeomorphism constraint by reexpressing 
the diffeomorphism invariant states using some intriguing formal 
expressions that will lead us to dealing with the Hamiltonian 
constraint.

Although we noticed that $\mathcal{H}_{diff}$ is not a subspace of $\mathcal{H}$,
there exists nevertheless a \lq \lq projection operator'' $\Pi$\footnote{Notice, that since
$\mathcal{H}_{diff}$ is not a subspace of $\mathcal{H}$, $\Pi$ is not really a projector
in the true sense of the word.},  
\begin{equation}
  \Pi \; : \; \mathcal{H} \rightarrow \mathcal{H}_{diff} \;.
\end{equation} 
It acts as
\begin{equation}
\label{projectorons}
  \Pi |S \rangle = |s \rangle \;,
\end{equation} 
i.e. a spin network state in $\mathcal{H}$ is mapped to the corresponding diffeomorphism 
invariant equivalence class in $\mathcal{H}_{diff}$.
In the following, we are going to describe its construction.
We start by formally defining a measure on $Diff(M)$, which is required to satisfy
\begin{equation}
\label{diffmeasure}
  \int_{Diff} [d\phi] = 1 \;,
\end{equation}
and
\begin{equation}
\label{deltas}
  \int_{Diff} [d\phi] \, \delta_{S, \phi \cdot S} = c(s) \;.
\end{equation}
Loosely speaking, (\ref{deltas}) refers to taking an embedded spin network,
acting with diffeomorphisms on it (i.e. displace it smoothly in the manifold), and finally 
moving it back to the spin network one started with. Then $c(s)$ just counts the number of ways
one can do this.

Now, a diffeomorphism invariant knot state $|s \rangle$ can be written as
\begin{equation}
\label{sasint}
  |s \rangle = \int_{Diff} [d\phi] \, |\phi \cdot S \rangle \;, \hspace{1cm} S \in s \;.
\end{equation}
Using only the definitions (\ref{diffmeasure})--(\ref{sasint}), one can in fact derive
(\ref{s_scalar1}) and (\ref{s_scalar2}).
Furthermore, we can also give a more explicit (but still formal) expression of the
projection operator $\Pi$. First, note that the generator of
the diffeomorphism constraint $\vec{D}[\vec{f}]$, which is a smooth vector field 
$\vec{f}$ on $M$, is an element of the Lie algebra of $Diff(M)$.
Then, from (\ref{actons}) and (\ref{sasint}) one concludes
\begin{equation}
  |s \rangle = \int_{Diff} [d\phi] \, U(\phi) | S \rangle
        = \int [d\vec{f}]\, e^{i \vec{f} \vec{D}}\, | S \rangle \;,
\end{equation}
where in the second step we have expressed a group element $U(\phi) \in U(Diff)$ as the 
exponential of an element of the Lie algebra, and formally integrated over the algebra
rather that the group.
From this and (\ref{projectorons}) we can immediately read off the projector
\begin{equation}
\label{projector_ex1}
  \Pi = \int [d\vec{f}]\, e^{i \vec{f} \vec{D}} \,.
\end{equation}
Finally, we also obtain a diffeomorphism invariant 
quadratic form\footnote{Notice that the quadratic form 
$\langle {\ } | {\ } \rangle_{diff}$ is highly degenerate.} on $\mathcal{H}$ via
\begin{equation}
\label{physquadratform}
  \langle S | S' \rangle_{diff} \equiv \langle S | \Pi | S' \rangle
        = \int [d\vec{f}]\, \langle S | e^{i \vec{f} \vec{D}} | S' \rangle \;,
\end{equation}
where one should notice that the spin networks $S$ and $S'$ are itselves not 
diffeomorphism invariant. Hence it follows (roughly) that the knowledge
of the \lq \lq matrix elements'' $\langle S | \Pi | S' \rangle$ of the projection operator 
is equivalent to the solution of the diffeomorphism constraint!

\subsection{The Hamiltonian Constraint, Spin Foam and Physical Observables}
\label{sect_hamiltonian_constraint}
In Fig.~\ref{construction_of_H} we illustrated the general outline for a step by 
step construction of 
the physical Hilbert space by solving the quantum constraint operators one after another.
Carrying this out, we were led from the unconstrained Hilbert space $\mathcal{H}$
firstly to the gauge invariant space 
$\mathcal{H}_0$, equipped with an orthonormal basis of spin network states 
$|S \rangle$, and secondly, as we described in sect.~\ref{diffsagain}, to the 
diffeomorphism invariant Hilbert space $\mathcal{H}_{diff}$, for which it was also possible to
define an orthonormal basis $|s \rangle$ of $s$-knot states. 
The final step, marked with a question mark in Fig.~\ref{construction_of_H},
remains to be done: the physical states of the theory should lie in the kernel of the quantum 
Hamiltonian constraint operator. Of 
course, we do not expect to find a complete solution of the 
Hamiltonian constraint, which would correspond to a complete 
solution of the theory.  Rather, we need a well posed 
definition of the Hamiltonian constraint, and a strategy to 
compute with it and to unravel its physical content.

Here, we will give only a sketchy account of the definition of 
the Hamiltonian constraint.  On the other hand, we will 
illustrate the way of using this constraint a bit more in 
detail.  The idea we will illustrate is to search the solution 
of the constraint by constructing a projector on physical 
states, using the procedure we described in the last section
for the diffeomorphism constraint as a guide to the solution of the Hamiltonian 
constraint. This construction will lead us 
to the so-called \emph{spin foam models}, which represent a covariant 
formulation of the dynamics of quantum gravity.
In the light of recent developments, these models provide
the most exciting and promising approach to the subject. 

\subsubsection{A Simple Example.}
We have already considered the construction of a projection operator in relation to 
the diffeomorphism constraint in sect.~\ref{diffsagain}. Nevertheless, it is instructive
to give here a simple toy example that should explain the procedure in a more 
accurate way.

Consider a simple dynamical quantum mechanical system with an unconstrained Hil\-bert 
space of square integrable functions over $\mathbb{R}^2$, 
i.e. $\mathcal{H} = L^2(\mathbb{R}^2)$.
Let the system be constrained by demanding invariance with respect to rotations around the
$z$-axis, i.e by having the angular momentum operator 
\begin{eqnarray}
\label{angmom1}
\hat{J} &:=& \hat{J}_z = i\,(x\, \partial_y - y\, \partial_x) \\
        \label{angmom2}
        &\widehat{=}& \hat{J}_{\varphi} = i\, \partial_{\varphi} 
\end{eqnarray}
as the quantum constraint.
In (\ref{angmom1}) and (\ref{angmom2}) we considered two representations, namely 
the cartesian and the polar coordinate one, in which the 
wave functions appear as $\Psi(x,y)$ or 
$\Psi(r,\varphi)$, respectively. We will confine ourselves to 
the latter one.
The physical Hilbert space $\mathcal{H}_{phys}$ is given as the subspace of $\mathcal{H}$
subject to
\begin{equation}
\label{Jconstraint}
 \hat{J} \Psi = 0 \; ,
\end{equation}
i.e. the physical state functionals are required to lie in the kernel of the 
quantum constraint operator $\hat{J}$.
We know that $\hat{J}$ is the generator of the group $U(1)$ with parameter $\alpha$, 
acting as
\begin{eqnarray}
  U(1) \times \mathbb{R}^2  &\rightarrow& \mathbb{R}^2 \\
  \label{actiononphi}
  \big( \alpha, (r, \varphi )\big) &\mapsto& ( r, \varphi + \alpha ) \;.
\end{eqnarray}
Due to compactness of $U(1)$, the constraint equation (\ref{Jconstraint}) 
could be solved directly. However, we will follow a different path. We try to solve 
the problem using a projection operator
\begin{equation}
  \Pi \; : \hspace{0.5cm} \mathcal{H} \rightarrow \mathcal{H}_{phys} \;.
\end{equation}
Note that the (finite) action of the constrait on a general state functional
$\Psi(r, \varphi)$ is given by
\begin{equation}
\label{finiteaction}
e^{i \alpha \hat{J}}\, \Psi(r, \varphi) = \Psi(r, \varphi + \alpha) \;.
\end{equation}
Hence, the projection operator $\Pi$ on physical states is defined as
\begin{equation}
\label{projector_ex2}
  \Pi = \frac{1}{2 \pi} \int_0^{2\pi} d\alpha \, e^{i \alpha \hat{J}} \;.
\end{equation}
It acts on $\Psi(r,\varphi) \in \mathcal{H}$ as follows,
\begin{eqnarray}
  \Pi \,\Psi(r,\varphi) &=& \frac{1}{2 \pi} \int_0^{2\pi} d\alpha \, e^{i \alpha \hat{J}} \Psi(r,\varphi) \\
        &=& \frac{1}{2 \pi} \int_0^{2\pi} d\alpha \, \Psi(r,\varphi + \alpha) = \tilde{\Psi} (r) \;,
\end{eqnarray}
resulting in a new function $\tilde{\Psi}(r)$ which is independent of $\varphi$, just as one  
might have expected for the physical states. Applying $\Pi$ a second time proves the correct 
projector property $\Pi^2 = \Pi$. However, because of the existence of the projector, 
there is no need to perform calculations in the 
physical subspace $\mathcal{H}_{phys}$, rather one can stay in the unconstrained
Hilbert space $\mathcal{H}$---a remarkable simplification!
Using the scalar product
\begin{equation}
  \langle \Psi | \Phi \rangle = \int_0^{2\pi} \!\!\int_0^{\infty}\! d\varphi\, dr 
        \: \overline{\Psi(r,\varphi)} \, \Phi(r,\varphi) 
\end{equation}
in $\mathcal{H}$, one arrives at the important result
\begin{equation}
\label{matrixelements}
  \langle \Psi | \Phi \rangle_{phys} \equiv \langle \Psi | \Pi | \Phi \rangle \;.
\end{equation}
This equation is similar to the result obtained in the discussion of the 
diffeo\-mor\-phism constraint in sect.~\ref{diffsagain}.
The quadratic form $\langle \, | \, \rangle_{phys}$
in $\mathcal{H}_{phys}$ is indeed expressed as a scalar product over states
which lie in $\mathcal{H}$. Thus, knowing the matrix elements (\ref{matrixelements})
of the projection operator in the unconstrained Hilbert
space is equivalent to having solved the constraint!

It is worth mentioning, that a similar scheme can be applied to operators.
Suppose there exists a non gauge invariant\footnote{For later convenience, we refer to
the symmetry in this example as a \lq \lq gauge'' symmetry.}
operator $O = O(r, \varphi)$ on $\mathcal{H}$. Then  a
fully gauge invariant operator $R=R(r)$ in $\mathcal{H}_{phys}$ can be 
constructed by defining
\begin{equation}
\label{projonobs}
  R := \Pi \,O\, \Pi \;.
\end{equation}
The calculation of matrix elements of the physical operator $R$ is then reduced to
a calculation in the unconstrained Hilbert space, which gives 
\begin{equation}
  \langle \Psi | O | \Phi \rangle_{phys} \equiv \langle \Psi | \Pi \,O\, \Pi | \Phi \rangle \;. 
\end{equation}

\subsubsection{The Hamiltonian Constraint.}
Let us now proceed with the application of the projector method to the Hamiltonian
constraint in quantum gravity, the only and least understood constraint which still remains 
to be solved in order to describe the dynamics of the theory. 
The following discussion is mainly based on plausibility considerations
resulting in rough arguments. A more complete treatment would include 
exponentially increasing efforts, which lies beyond the scope of this 
lecture.

The Lorentzian Hamiltonian constraint can be written as the sum of two terms, where one is the 
Euclidean Hamiltonian constraint.
Here, we deal for simplicity only with this Euclidean part, which means that we actually 
consider Euclidean gravity only.
Classically, the constraint reads roughly
\begin{equation}
\label{hamconstraint}
  H_{cl} \simeq F_{ab} E^a E^b + \mbox{Lorentzian part} \;.
\end{equation}
$E^{a,b}$ are the triads, and $F_{ab}$ is the curvature of the
connection, which, we recall, is an antisymmetric tensor.

When passing over to the quantum theory, $H_{cl}$ is promoted to an operator which
has to be suitably regularized. 
A typical regularization process consists of the following steps.
First, introduce a regularization parameter $\epsilon$,
and replace the classical expression (\ref{hamconstraint}) with a 
regularized, $\epsilon$ dependent one, written in terms of quantities that 
we know how to promote to quantum operators, and which tend to 
$H_{cl}$ as $\epsilon$ tends to zero. In particular, $F$ is replaced by the
holonomy of an $\epsilon-$size loop. In the second step, 
replace the classical quantities with their quantum 
analogues leading to the Hamiltonian operator 
$\hat{H}_{\epsilon}$. Finally, the parameter is 
forced to go to zero, $\epsilon \rightarrow 0$, yielding a 
well-defined quantum Hamiltonian operator 
$\hat{H}_{\epsilon} \rightarrow \hat{H}$.

We will not carry out this construction explicitely but only mention that there
exist \emph{several} different versions. The first completely
consistent construction, yielding a well-defined and finite operator, was obtained
by Thiemann in \cite{Thiemann96b}.

However, the key point which is common to all different regularization procedures 
is the vanishing of the action of the Hamiltonian operator on the
holonomy $U[A, \gamma]$. That is
\begin{equation}
\label{HonU}
\hat{H}(x) \, U[A, \gamma] = 0\;,
\end{equation}
if $x$ is on an interior point of the curve $\gamma$.
The reason for this can roughly be understood as follows.
If we replace the triad in (\ref{hamconstraint}) with its quantum analogue and apply
the resulting operator to the holonomy (without bothering about regularization), we obtain
\begin{equation}
\label{H_on_holonomy}
  F_{ab} \,\frac{\delta}{\delta A_a} \frac{\delta}{\delta A_b} \, U[A, \gamma] 
        \; \sim \; F_{ab} \, \dot{\gamma}^a \dot{\gamma}^b = 0 \;.
\end{equation}
While $F_{ab}$ is an antisymmetric tensor, the product of $\dot{\gamma}^{a}$ and 
$\dot{\gamma}^{b}$, which are tangent to $\gamma$, is, on the other hand, symmetric.
Thus the result is zero, since we contract an antisymmetric tensor with a 
symmetric quantity.  This derivation has only a formal character since 
an infinite coefficient multiplies the right hand side of 
(\ref{H_on_holonomy}). Anyhow, a careful calculation using rigorous regularization 
yields the same result. 

However, calculating the action of $\hat{H}$ on spin 
network states, the result turns out to be not equal to zero,
\begin{equation}
  \hat{H}(x) \, \Psi_S \neq 0 \;.
\end{equation}
This fact is due to the end points of the links, i.e. the nodes of the spin
network. At a node,
the tangent vectors $\dot{\gamma}^a$ and $\dot{\gamma}^b$ in (\ref{H_on_holonomy}) might
refer to distinct links adjacent to the node. This results in 
terms with non-zero contributions. Hence one concludes that the Hamiltonian 
constraint operator acts on the \emph{nodes} only.

\begin{figure}
\centerline{\includegraphics[width=8cm]{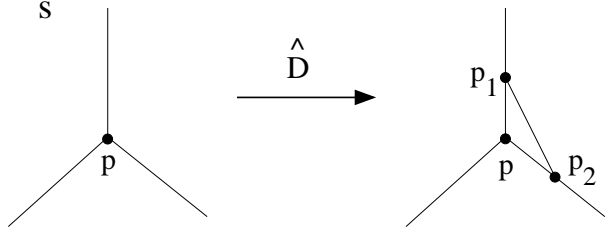}}
\caption[]{The action of the Hamiltonian constraint on a trivalent node.}
\label{hamonvertex}
\end{figure}

The result of the action of $\hat{H}$ on a spin network
state $| S \rangle$ turns out to be given by 
\begin{equation}
\label{H_on_s}
  \hat{H}[N] \, | S \rangle = 
	\sum_{\mbox{\scriptsize nodes}\, n\,  \mbox{\scriptsize of}\, S}
  	A_n \, N(x_n)\, \hat{D}_n \, | S \rangle \;,
\end{equation}
where $x_n$ refers to the point in which the node $n$ is located.
The action of the operator $\hat{D}_n$ is illustrated in 
Fig.~\ref{hamonvertex}. 
An extra link with color one connecting two points $p_1$ and $p_2$ lying on
distinct links adjacent to the node $p$ is created when acting on a single node.
The color of the link between $p$ and $p_1$, as well as between $p$ and $p_2$ 
is altered and the state is multiplied by a coefficient $A$.
Explicit expressions are computed in \cite{Borissov97}.
Moreover, $\hat{H}[N]$ is the Hamiltonian constraint smeared
with a scalar function $N(x)$ given by
\begin{equation}
  \hat{H}[N] = \int d^3x \, N(x) \,\hat{H}(x) \;.
\end{equation}

\subsubsection{Spin Foam.}
Now we want to define the physical Hilbert space $\mathcal{H}_{phys}$ 
using the projector
method explained above, starting from the diffeomorphism invariant Hilbert space 
$\mathcal{H}_{diff}$ by considering $s$-knot states. 
Similar to (\ref{projector_ex1}) or (\ref{projector_ex2}), respectively, we construct the 
projection operator 
\begin{equation}
\label{P}
   P = \int [dN] \, e^{i \hat{H}[N]} =  \int [dN] \, e^{i \int N \, \hat{H}} \;.
\end{equation}
In the abstract spin network basis, the matrix elements of $P$ are
\begin{equation}
\label{matrixelofP}
  \langle s | P | s' \rangle 
  =  \langle s | \int [dN] \, e^{i \int N \, \hat{H}} | s' \rangle \;.
\end{equation}
It can be shown, that a diffeomorphism invariant notion of integration exists for 
this functional integral \cite{Rovelli98}.
According to (\ref{physquadratform}) or (\ref{matrixelements}), respectively, we use the 
matrix elements of the projector to define the quadratic form 
\begin{equation}
\label{quadraform}
  \langle s | s' \rangle_{phys} = \langle s | P | s' \rangle\;. 
\end{equation}
The physical Hilbert space $\mathcal{H}_{phys}$ is then defined over $\mathcal{H}_{diff}$, from 
which we started, via this quadratic form.

In order to calculate the matrix elements (\ref{matrixelofP}) of the projector,
the exponent is expanded. Neglecting many technicalities which are given in 
\cite{ReisenbergerRovelli97} the expansion looks schematically as follows,
\begin{equation}
  \langle s | P | s' \rangle \sim \langle s | s' \rangle +
        \int [dN] \, \left( \langle s | \hat{H} | s' \rangle\ + 
        \langle s |  \hat{H} \hat{H} | s' \rangle\ + \ldots \right) \;.
\end{equation}
Using now the action (\ref{H_on_s}) of $\hat{H}$ on spin network states, we obtain
\begin{equation} 
\label{expansion}
  \langle s | s' \rangle_{phys} = \langle s | P | s' \rangle
        \sim  \langle s | s' \rangle +  
        \sum_{\mbox{\scriptsize nodes}\, n\,  \mbox{\scriptsize of}\, s'}
A_n \, \langle s | s'_n \rangle + \ldots \;,
\end{equation}
where we \lq \lq integrated out'' integrals of the type 
\begin{equation}
  \int [dN] \,  \big( N(x_1) \cdots N(x_n) \big) \;.
\end{equation}
Equation (\ref{expansion}) admits an extremely compelling graphical interpretation
as a sum over histories of evolutions of $s$-knot states. 
This reveals the meaning of the projector as a propagator in accordance with Feynman.

\begin{figure}[t]
\centerline{\includegraphics[width=5.5cm]{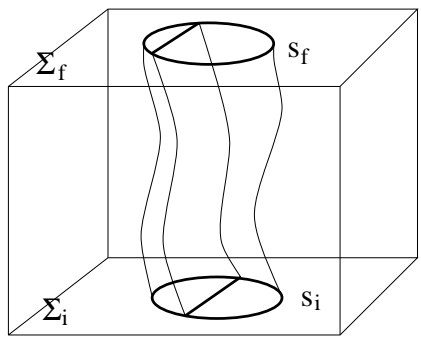}}
\caption[]{The diagram corresponding to a term of order zero.}
\label{order0}
\end{figure}

To see this more clearly, consider the 4-manifold $\mathcal{M} = \Sigma \times [0,1]$. 
The hypersurfaces at the boundary of $\mathcal{M}$, corresponding to the values
$0$ and $1$ in the interval, are denoted as $\Sigma_i$ and $\Sigma_f$, respectively.
We define the \lq \lq initial state'' on $\Sigma_i$ as $s_i := s'$, and the 
\lq \lq final state'' on $\Sigma_f$ as $s_f := s$.
Then the term $\langle s_f | s_i \rangle$, which is of order zero in the expansion 
(\ref{expansion}), is non-vanishing only if $s_f = s_i$. In other words, the corresponding 
graphs have to be continuously
deformable into each other such that the colors of the links and nodes match.
Graphically, this is expressed by sweeping
out a surface $\sigma = \sigma_i \times [0,1]$, as shown in Fig.~\ref{order0}. 
The surface is formed by 2-dimensional submanifolds of $\mathcal{M}$---so-called 
faces---which 
join in edges. The faces are swept out by spin network links, and the edges by
the nodes. Thus every face of $\sigma$ is colored just as the underlying link, and to every  
edge the intertwiner of the underlying node is associated.

\begin{figure}[b]
\centerline{\includegraphics[width=5.5cm]{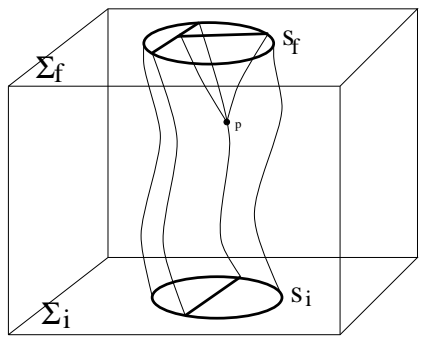}}
\caption[]{A first order diagram.}
\label{order1}
\end{figure}

Next, we consider a first order term $\langle s | s'_n \rangle$ in the expansion 
(\ref{expansion}). These term appear as the result of a single action of 
the Hamiltonian constraint, i.e. they correspond to
adding (or removing) one link, or, equivalentely, two nodes into a spin network, cf.
Fig.~\ref{hamonvertex}.
The situation is similar to the one described for the term of order zero, but now
at some point $p$ of $\sigma$ the surface branches as shown in Fig.~\ref{order1}.
Thus the graph of $s_i$ is not equal to the one which is 
associated to $s_f$ any more, as opposed to the previous case. 
The surfaces are again colored corresponding to the underlying links.

\begin{figure}[t]
\centerline{\includegraphics[width=3.5cm]{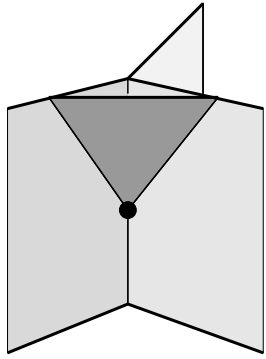}}
\caption[]{The elementary vertex.}
\label{vertex}
\end{figure}

The picture one should have in mind is the following. $\mathcal{M}$ can be imagined as
a space--time, and $s_i$ is a spin network that evolves continuously in a coordinate 
denoted as \lq \lq time'' up to a point $p$ where the spin network branches
because of the action of the Hamiltonian constraint. At a branching point
the single node $p_i$ degenerates in the sense of being transformed into three nodes, 
each distinct pair being connected by a link. 
The accompanying branching of the surface in $p$ is called the 
\emph{elementary vertex} of the theory. It is the simplest geometric 
vertex, see Fig.~\ref{vertex}.

\begin{figure}[h]
\centerline{\includegraphics[width=8.5cm]{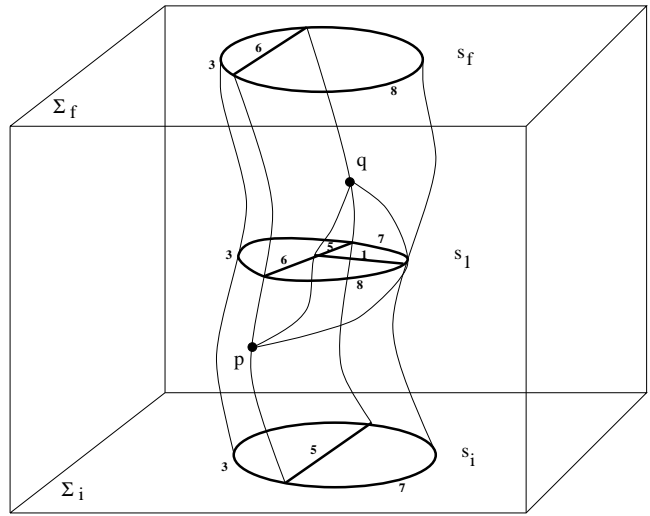}}
\caption[]{A term of second order.}
\label{order2}
\end{figure}

Finally, we will have a closer look to a second order term. In this example we give 
the coloring of the surfaces
as explicitely shown in Fig.~\ref{order2}. 
\vspace*{0.2cm}
Consider the transition from
an (abstract) spin network $s_i$ with two trivalent nodes connected by three links with colors 
$(3, 5, 7)$ to the $s$-knot $s_f$ with the same underlying graph but different coloring 
$(3, 6, 8)$. There an intermediate state with four nodes emerges, such that an 
elementary creation 
as well as an elementary annihilation vertex occur giving rise to the coloring specified in 
Fig.~\ref{order2}.
Although our considerations are extremely simplified, it is nevertheless plausible 
to write the expansion (\ref{expansion}) of $\langle s | P | s' \rangle$
as a sum over topologically inequivalent branched colored surfaces 
$\sigma$ bounded by $s_i$ and $s_f$. These surfaces are 
called \emph{spin foams} \cite{ReisenbergerRovelli97,Baez98}. 
Each surface $\sigma$ represents the history of the initial $s$-knot state. It is weighted
by the product of coefficients $A_{\nu}$, which are associated to the vertices of $\sigma$.
Recall that these coefficients showed up in (\ref{H_on_s}) as a result of the action of
the Hamiltonian constraint on $s$-knot states. They depend only on the coloring
of the faces and edges adjacent to the relevant vertices.
In the end, we obtain 
\begin{equation}
\label{sumoverfoams}
  \langle s | s' \rangle_{phys} = \langle s | P | s' \rangle = 
        \sum_{\mbox{\scriptsize spin} \atop \mbox{\scriptsize foams}\, \scriptstyle \sigma} 
        \prod_{\mbox{\scriptsize vertices} \atop  \scriptstyle \nu \in \sigma} A_{\nu} \;
\end{equation}
for the transition amplitude between the two (abstract) spin network states $s_i$ and $s_f$.
It is designated as a transition amplitude because of the obvious formal analogy to the 
expressions in standard quantum field theory,
giving rise to the interpretation of Fig. \ref{order0}, \ref{order1} and \ref{order2}
roughly as \lq \lq Feynman diagrams'' of quantum gravity.

This interpretation is reinforced by a number of independent results.  
For instance, certain discretized covariant approaches to quantum 
gravity lead precisely to a \lq \lq sum over discretized 4-geometries'', very 
similar to (\ref{sumoverfoams}), cf. \cite{Reisenberger97}.
Inspired by the construction above, Baez 
has defined a general notion of {\em spin foam model} and studied 
the structure of these models in general. See \cite{Baez98,Baez98b} and 
references therein. 

\subsubsection{Physical Observables.}
To round off this section, we briefly comment on an application of the projector method 
to the calculation of physical observables. 
As before, we will not consider problems that arise with 
normalization, or ill-defined expressions that might occur,
but rather concentrate on the conceptual framework. For a more
detailed account to this subject we refer to \cite{Rovelli98}.

Trying to construct physical observables, i.e. self-adjoint operators which are invariant 
under the 4-dimensional diffeomorphism group is a well-known difficulty in 
(quantum) general relativity.
Instead, as explained in the last sections, we can immediately define a fully gauge invariant
observable by starting from an operator $O$ acting on $\mathcal{H}_{diff}$,
which is invariant under 3-dimensional diffeomorphisms.
Using what we noticed in (\ref{projonobs}), the fully gauge invariant operator $R$ is given by
\begin{equation}
  R = P \,O\, P \;.
\end{equation}
The projector $P$ onto the physical Hilbert space is defined in (\ref{P}).
Indeed, $R$ is invariant under 4-dimensional diffeomorphisms.
Thus, the expectation value in a physical state is 
\begin{equation}
\label{matrelementop}
  \langle s | O | s \rangle_{phys} := \langle s | P \,O\, P | s \rangle \;.
\end{equation}
Doing a similar manipulation of the matrix elements (\ref{matrelementop}) as
above, we obtain the expression for the expectation values in the spin foam version as 
\begin{equation}
\label{foamyexpectationval}
  \langle s | O | s\rangle_{phys} \; \sim 
        \sum_{\mbox{\scriptsize spin} \atop \mbox{\scriptsize foams}\, \scriptstyle \sigma}
        \left( \sum_{\tilde{s}} O (\tilde{s}) \right) 
        \prod_{\mbox{\scriptsize vertices} \atop  \scriptstyle \nu \in \sigma} A_{\nu} \;.
\end{equation}
For simplicity, we have chosen $O$ to be diagonal. Furthermore, $\tilde{s}$ are all 
possible spin networks that cut a spin foam $\sigma$ (i.e. a branched colored 2-surface) 
into two parts, a future and a past one. These $\tilde{s}$ may be considered 
as (ADM-like) spatial slices that cut a given spin foam.

A closer examination of (\ref{foamyexpectationval}) reveals that the first 
summation has to be performed over all possible spin 
foams $\sigma$. On top of that, for each of these spin foams, all of its
spatial slices have to be summed up.
Without going into details, we briefly mention its appealing geometrical
interpretation as an \lq \lq integration over space--time'', or more
precisely, as an \lq \lq integration'' over the location of the ADM surfaces in
(the quantum version of classical) 4-dimensional space--time.
Thus, expectation values of physical observables are given as averages over the spin foam 
in an intuitively similar manner as one is used to from standard quantum field theory.
Moreover, this method provides a framework for the non-perturbative, 
space--time covariant formulation of a diffeomorphism invariant quantum field theory.

However, so far we haven't mentioned any problems that arise. Recall first,
that we considered only the Euclidean part of the Hamiltonian constraint.
Furthermore, it is still unclear what shape physical observables $O$, which are at least
required to yield finite results, should take.
Intuitively, we might expect that observables of the form 
\begin{equation}
  O = \tilde{O} \times \delta(\mbox{\footnotesize something}) \;,
\end{equation}
might be finite, and might correspond to the realistic relational 
observables discussed above. 

But so far the problem of finding physical observables in quantum gravity is still very 
little explored territory, and our considerations may at best give some 
vague ideas of what remains to be done.

%%%%%%%%%%%%%%%%%%%%%%%%%%%%%%%%%%%%%%%%%%%%%%%%%%%%%%%%%%%%%%%%%%%%%%%%%%%%%%%%%%%%%%%
%%%%%%%%%                              Section 6                              &&&&&&&&&
%%%%%%%%%%%%%%%%%%%%%%%%%%%%%%%%%%%%%%%%%%%%%%%%%%%%%%%%%%%%%%%%%%%%%%%%%%%%%%%%%%%%%%%

\section{Open Problems and Future Perspectives}
\label{sect_perspectives}

This series of lectures was devoted to loop quantum gravity, a non-per\-tur\-ba\-tive canonical
formulation for a quantum theory of gravitation. We introduced the basic principles
of the theory in the kinematical regime, including spin network states which provide 
an orthonormal basis in the gauge invariant Hilbert space.
As an application, one of the most exiting results obtained in the last few years,
the discreteness of geometry, was examined by considering the quantization of the area.
Furthermore, by taking the basic principles of general relativity seriously, we have shown 
by discussing the topics of diffeomorphism invariance and observability in general relativity,
that loop quantum gravity is well-adapted for a quantum theory of gravitation.

Finally, in order to examine also the non-perturbative \emph{dynamics} of quantum 
gravity a little, an ansatz for the construction of the physical Hilbert space by 
means of a projection method was explained.
We tried to clarify its interpretation in terms of a spin foam model, in which the projection 
operator itself plays the role of a propagator for the space--time evolution 
of (abstract) spin networks.  Its Feynman diagram like graphic representation was 
presented as well. We also gave the prospects for a possible calculation of expectation values 
of operators representing physical observables, by using the spin foam formalism.

There are several open questions which remain to be explored. 
We mentioned that, because of different regularization schemes, 
there exist several versions of the Hamiltonian constraint. Thus, one of
the most intriguing questions would certainly be to find the 
\lq \lq right'' consistent Hamiltonian constraint, i.e. the one which has the 
correct classical limit.
Closely related is the question of how such a classical limit should be studied. 
What are the coherent states? What is the ground state of the theory?
Does a notion of \lq \lq ground state'' make sense at all, in a general covariant theory?

The problem of constructing 4-dimensional diffeomorphism invariant 
observables is crucial.  We do know many 4-dimensional 
diffeomorphism invariant observables in general relativity: in fact, 
we use them in the classical applications of general relativity, which 
are nowdays extremely numerous.  But to express such observables in the 
quantum theory is still technically hard.  In particular, in order to 
compare loop quantum gravity with particle physics approaches, and to 
make contact with traditional quantum field theory, it would be 
extremely useful to be able to compute scattering amplitudes in an 
asymptotically flat context.  Some kind of perturbation expansion 
should be used for such a project.  But in this context the notion of 
\lq \lq expansion'', and \lq \lq perturbative'' are delicate (expand 
around what?).  For these problems, the spin foam formalism may turn 
out to be essential, since it provides a space--time formulation of a 
diffeomorphism invariant theory. 

We close these lectures by expressing the wish that some of the students 
that so enthusistically attended them will be the ones able to solve 
these problems, to give us a fully convincing quantum theory of 
space--time, and thus push forward this extraordinary beautiful 
adventure, which is exploring Nature and its marvellous and 
disconcerting secrets.

%%%%%%%%%%%%%%%%%%%%%%%%%%%%%%%%%%%%%%%%%%%%%%%%%%%%%%%%%%%%%%%%%%%%%%%%%%%%%%%%%%%%%%%
%%%%%%%%%                              References                             &&&&&&&&&
%%%%%%%%%%%%%%%%%%%%%%%%%%%%%%%%%%%%%%%%%%%%%%%%%%%%%%%%%%%%%%%%%%%%%%%%%%%%%%%%%%%%%%%

%\nocite{*}

\end{document}